\documentclass[a4paper,12pt]{article}
\usepackage{jheppub} 
\usepackage{latexsym,graphicx,amssymb,color,amsmath}
\usepackage{relsize,ccaption,url,latexsym,epsfig,a4wide}
\usepackage{pdfsync}
\usepackage[numbers,sort&compress]{natbib}
\usepackage{hyperref}

\usepackage[table]{xcolor}

\newcommand{\RR}{\mathbb{R}} 
\newcommand{\QQ}{\mathbb{Q}} 
 
\newcommand{\HH}{\mathbb{H}} 
\newcommand{\ZZ}{\mathbb{Z}}

\newcommand{\Hh}{\mathcal{H}}

\newcommand{\M}{\mathcal{M}}
\newcommand{\N}{\mathcal{N}}
\newcommand{\calO}{\mathcal{O}}
\newcommand{\G}{\mathcal{G}}
\newcommand{\D}{\mathcal{D}}

\newcommand{\T}{\mathcal{T}}
\newcommand{\R}{\mathcal{R}}
\renewcommand{\L}{\mathcal{L}}
\newcommand{\F}{\mathcal{F}}
\newcommand{\I}{\mathcal{I}}
\newcommand{\C}{\mathcal{C}}

\newcommand{\qu}{\bar}
\newcommand{\bbar}{|\!|}
\newcommand{\ii}{\mathbf{i}}
\newcommand{\jj}{\mathbf{j}}
\newcommand{\kk}{\mathbf{k}}
\newcommand{\om}{\boldsymbol{\omega}}
\newcommand{\iiO}{\boldsymbol{\iota}_O}
\newcommand{\iiI}{\boldsymbol{\iota}_I}
\newcommand{\eei}{\mathbf{e}}

\newcommand{\rrangle}{\rangle\!\!\!\;\rangle}
\newcommand{\llangle}{\langle\!\!\!\;\langle}
\newcommand\nop[1]{\mathop{:\!#1\!\!:}}
\DeclareMathOperator{\spa}{span}

\newtheorem{theorem}{Theorem}

\DeclareMathOperator{\Tr}{Tr}

\DeclareMathOperator{\rk}{rk}
\DeclareMathOperator{\Spin}{Spin}

\numberwithin{equation}{section}

\def\beq{\begin{equation}}
\def\ee{\end{equation}}

\allowdisplaybreaks[3]

\captionnamefont{\bfseries}
\title{
On symmetries of $\N=(4,4)$ sigma models on $T^4$}

\author{Roberto Volpato}

\affiliation{Max-Planck-Institut f\"ur Gravitationsphysik, Am M\"uhlenberg\,1, D-14476 Golm, Germany}

\emailAdd{roberto.volpato@aei.mpg.de}

\abstract{
Motivated by an analogous result for K3 models, we classify all groups of symmetries of non-linear sigma models on a torus $T^4$ that preserve the $\N=(4,4)$ superconformal algebra. 
The resulting symmetry groups are isomorphic to certain subgroups of the Weyl group of $E_8$, that plays a role similar to the Conway group for the case of K3 models. Our analysis heavily relies on the triality automorphism of the T-duality group $SO(4,4,\ZZ)$. As a byproduct of our results, we discover new explicit descriptions of K3 models as asymmetric orbifolds of torus CFTs.}

\begin{document}

\begin{flushright} AEI-2014-005\end{flushright}

\maketitle

\section{Introduction and summary}

Non-linear sigma models on  tori provide the basic examples of string theory compactifications. The study of the discrete groups of symmetries and the orbifolds of supersymmetric torus models, both in heterotic or type II superstring theory, have been the subject of intensive study since the eighties \cite{Dixon:1985jw,Dixon:1986jc,Ibanez:1987pj} (see also \cite{Reffert:2006du} and references therein). These models represent both a fruitful arena for the formal understanding of string compactifications and are of great interest from a phenomenological viewpoint.

In this paper,  we focus on type II non-linear sigma models with target space four dimensional tori $T^4$ and classify the groups of symmetries that commute with the (`small') world-sheet $\N=(4,4)$ superconformal algebra. This analysis is mainly motivated by the relation between such torus models and non-linear sigma models on K3. There are few K3 models for which an explicit and complete description as conformal field theories is available; among them, torus orbifolds play a fundamental role. The interest in K3 superstring compactifications has received new impetus in the last few years with the discovery of the Mathieu moonshine phenomenon. This conjecture, originated from an observation of Eguchi, Ooguri and Tachikawa (EOT) in  \cite{eot10}, proposes a connection between the elliptic genus of K3 and a finite sporadic simple group, the Mathieu group $M_{24}$.  After the original EOT proposal, a considerable amount of evidence in favour of this conjecture has been compiled \cite{ch10,ghv10a,ghv10b,eghi11,ga12} and several different incarnations of the relationship between $M_{24}$ and various string compactifications on K3 have been uncovered \cite{Eguchi:2011aj,Cheng:2013kpa,Harvey:2013mda,Harrison:2013bya,Wrase:2014fja,
Gaberdiel:2012gf,Gaberdiel:2013nya,Cheng:2012tq,Cheng:2013wca}. Despite the amount of work on the subject, however, no satisfactory explanation of this phenomenon has been provided so far.

One of the most natural approaches towards an interpretation of Mathieu Moonshine is the analysis of the groups of discrete symmetries of K3 models. Orbifolds of non-linear sigma models on $T^4$ provide some of the simplest examples for such an investigation \cite{tawe11,tawe12,tawe12b}. In \cite{K3symm}, all possible groups of discrete symmetries of K3 models that commute with the $\N=(4,4)$ superconformal algebra have been classified. 
This result can be thought of as a stringy analogue of Mukai's theorem in algebraic geometry  \cite{Mukai,Kondo}, where the groups of symplectic automorphisms of K3 surfaces were considered. 
The rather surprising outcome of this classification is that all such groups of symmetries are subgroups of the Conway group $Co_0$, the group of automorphisms of the Leech lattice. Although $M_{24}$ itself is a subgroup of $Co_0$, it does not appear in the list of \cite{K3symm}. As a consequence, no K3 sigma model admits $M_{24}$ as its group of symmetries. It was observed in \cite{Gaberdiel:2012um} that, for  almost all K3 models, the symmetry group is actually a subgroup of $M_{24}$ and that most of the exceptions  correspond to torus orbifolds. The analysis of \cite{Gaberdiel:2012um} also showed the existence of a K3 model represented by an asymmetric orbifold of a torus model by a symmetry of order $5$.

This amount of work on non-linear sigma models on K3 led to a rather paradoxical situation: in fact, although non-linear sigma models on $T^4$ are much better understood than K3 models,  the classification of the groups of symmetries preserving the $\N=(4,4)$ superconformal algebra was known only for the latter. The goal of this paper is to fill in this gap. This completes the classification for all (known) models with $\N=(4,4)$ superconformal symmetry at central charge $c=\tilde c=6$, that are related to type II compactifications with (at least) $16$ space-time supersymmetries. Our main result is the following:

\medskip

{\sl The group $G$ of symmetries of a supersymmetric non-linear sigma model on $T^4$ that commutes with the (small) $\N=(4,4)$ superconformal algebra is
$$ G=\bigl( U(1)^4\times U(1)^4).G_0\ ,
$$ where $G_0$ is one of the following finite subgroups of $SU(2)\times SU(2)$:
\begin{itemize}
\item[--] Geometric groups: $\C_2,\quad \C_4,\quad \C_6,\quad  \D_8(2),\quad \D_{12}(3),\quad \T_{24}$
\item[--] Non-geometric groups: $\left\{\begin{array}{l} \T_{24}\times_{C_3}\T_{24},\qquad  \I_{120}\times_{\I_{120}}\bar\I_{120},\qquad \calO_{48}\times_{\calO_{48}} \bar\calO_{48}, \\[5pt] \D_{12}(3)\times_{\C_4} \D_{12}(3),\qquad  \D_{24}(6)\times_{\D_{24}(6)} \D_{24}(6),\\[5pt] \D_{8}(4)\times_{D_{4}(2)} \D_{8}(4)\end{array}\right.$ 
\end{itemize}
(Here, $\C_n$  denotes the cyclic group of order $n$, $\D_{4n}(n)$  the binary dihedral group of order $4n$ and $\T_{24}$, $\calO_{48}$, $\I_{120}$ the binary tetrahedral, octahedral and icosahedral groups, respectively. Finally, $\L\times_\F \R$ denotes a subgroup of $\L\times \R\subset SU(2)\times SU(2)$ of order $|\L||\R|/|\F|$. More details about the groups $G_0$ and the corresponding models are given in section \ref{s:groups} and appendix \ref{s:discrSpin}). 
}

\medskip

This rather abstract description of the symmetry groups deserves some further clarification (see section \ref{s:generalities} for details). The continuous normal subgroup $U(1)^4\times U(1)^4$ is generated by the zero modes of four holomorphic and four antiholomorphic $u(1)$ currents $\partial X^i$ and  $\bar\partial X^i$, and contains in particular the group $U(1)^4$ of translations along the four directions on the torus. 
The group $G_0=G/(U(1)^4\times U(1)^4)$ includes the rotations (fixed point automorphism) of the target space. If $G_0$ is generated only by automorphisms of the target space, we say that the group is (purely) geometric. In this case, the requirement that the superconformal algebra is preserved implies that $G_0$ is a subgroup of $SU(2)$. The properties of the purely geometric groups are well-known \cite{Wendland:2000ry} and their classification follows from a theorem by Fujiki \cite{Fujiki}.
In general, the group $G_0$ might contain elements that act asymmetrically on the left- and the right-moving fields (non-geometric group). In this case, $G_0$ is a subgroup of the product of two $SU(2)$, acting separately on the left- and right-moving fields. The $SU(2)$ subgroup of geometric symmetries is embedded diagonally in this general $SU(2)\times SU(2)$. Both in the geometric and non-geometric case, $G_0$ must act by automorphisms on the Narain lattice $\Gamma^{4,4}$. This condition constrains $G_0$ to be a discrete subgroup of $SU(2)\times SU(2)$ that depends on the specific torus model.

\bigskip

In every torus model, the group $G_0$ contains a central involution that flips the sign of all target space coordinates. For a generic model, this is the only  non-trivial element of $G_0$. The precise classification of all possible discrete subgroups $G_0$, or rather of the quotients $G_1=G_0/\ZZ_2$ by this central involution, is better understood by considering the action on the lattices of D-brane charges. In particular, as discussed in section \ref{s:RRDb}, the lattices $\Gamma^{4,4}_\textnormal{even}$ and $\Gamma^{4,4}_\textrm{odd}$ of even and odd dimensional D-brane charges, are both isomorphic to the winding-momentum (Narain) lattice. This is a peculiarity of non-linear sigma models on four dimensional tori and is closely related to the triality automorphism of the group $\Spin(4,4)$. The role of triality for non-linear sigma models on $T^4$ was emphasized in \cite{nawe01,KiObPio}.

\bigskip

In section \ref{s:groups}, using this triality, we obtain a useful and suggestive characterization of the groups $G_1$ as subgroups of $W^+(E_8)$, the group of even Weyl transformations of the $E_8$ root lattice. More precisely, the possible groups $G_1$ are precisely the subgroups $W^+(E_8)$  that fix pointwise a sublattice of $E_8$ of rank at least four. This result is the closest analogue of the classification theorem for K3 models: in this case, the possible groups of symmetries are exactly those subgroups of the Conway group $Co_0$ that fix pointwise a sublattice of the Leech lattice of rank at least four \cite{K3symm}. Whether the appearance of the Leech and of the $E_8$ lattice and their group of automorphisms is just an accident or if it has some physical meaning is still an open question.

\bigskip

Finally, in section \ref{s:orbifolds}, we discuss the orbifolds of torus models by cyclic groups of symmetries. By construction, these orbifolds, if consistent, must enjoy a $\N=(4,4)$ superconformal symmetry at central charge $c=\tilde c=6$, so that they are necessarily non-linear sigma models on $T^4$ or K3. It is easy to see that the orbifolds by cyclic subgroups of $U(1)^4\times U(1)^4$ always give rise to torus models. For each symmetry $g\in G$ with non-trivial image in $G_0$, we classify the nature (i.e., the topology of the target space) of the corresponding orbifold model. In this sense, it is useful to think  of $g$ as an element in a $\ZZ_2$ central extension $\ZZ_2.W^+(E_8)$ of the Weyl group of $E_8$. Indeed, the nature of the orbifold only depends on the conjugacy class of $g$ within $\ZZ_2.W^+(E_8)$. As a consistency check of our analysis, we verify that the classification of torus orbifolds matches perfectly with the results of \cite{Gaberdiel:2012um}, where a similar investigation was considered for orbifolds of K3 models.

\bigskip

Possible generalizations and applications of our results are discussed in section \ref{s:conclusions}. Some technical details about finite subgroups of $SU(2)\times SU(2)$ and integral lattices are relegated to the appendices.

\section{Generalities on torus models}\label{s:generalities}

In this section we review the main features of supersymmetric non-linear sigma models on $T^4$ and fix our notation and conventions.


Abstractly, a supersymmetric $T^4$-model is defined in terms of four left-moving $u(1)$ currents $$j^a(z)=i\partial \phi^a(z)$$ and four free fermions $\psi^a(z)$, $a=1,\ldots,4$, their right-moving analogues $\tilde \jmath^a(\bar z)=i\bar\partial \tilde\phi^a(\bar z),\tilde\psi^a(\bar z)$,  as well as `exponential fields' $V_\lambda(z,\bar z)$ labelled by the vectors $\lambda$ in an even unimodular lattice $\Gamma^{4,4}$ of signature $(4,4)$. 
The holomorphic fields have a mode expansion 
$$ j^a(z)=\sum_{n\in \ZZ} \alpha^a_n z^{-n-1}\ ,\qquad \psi^a(z)=\sum_{r\in \ZZ+\nu} \psi^a_r z^{-r-\frac{1}{2}}\ , 
$$ where $\nu=1/2$ and $\nu=0$ in the Neveu-Schwarz (NS) and Ramond (R) sectors, respectively, and the modes satisfy the (anti-)commutation relations
 $$[\alpha_m^a,\alpha_n^b]=m\delta^{ab}\delta_{m,-n}\ ,\qquad\qquad \{\psi_r^a,\psi_s^b\}=\delta^{ab}\delta_{r,-s}\ ,
$$
corresponding to the OPE
\beq\label{fundOPEs} j^a(z)j^b(w)\sim \frac{\delta^{ab}}{(z-w)^2}\ ,\qquad \psi^a(z)\psi^b(w)\sim \frac{\delta^{ab}}{(z-w)}\ .
\ee
 The anti-holomorphic fields $\tilde \jmath^a(\bar z),\tilde\psi^a(\bar z)$, $a=1,\ldots,4$, and their modes $\tilde\alpha^a_n,\tilde\psi^a_r$ satisfy analogous relations. The chiral algebra contains, in addition to the $u(1)^4$ `bosonic' algebra generated by $j^a$, $a=1,\ldots,4$,  also a `fermionic' $so(4)_1$ Kac-Moody algebra generated by $\nop{\psi^a(z)\psi^b(z)}$, $1\le a<b\le 4$, and $16$ fields of weight $(3/2,0)$ of the form $\nop{\psi^i(z) j^k(z)}$. In particular, any torus model contains a (small) $\N=(4,4)$ superconformal algebra with central charge $c=\tilde c=6$, that will be described in detail in section \ref{s:smallN4}.   The zero modes of the currents generate a continuous group of symmetries of the conformal field theory. In particular, we denote by
\beq\label{u1bos} U(1)^4_L\ ,\qquad\qquad U(1)^4_R\ ,
\ee the groups of symmetries generated by the zero modes $j^a_0$ and $\tilde j^a_0$ of the left- and right-moving currents and by
\beq\label{so4ferm} SO(4)_{f,L}\ ,\qquad\qquad SO(4)_{f,R}\ ,
\ee the groups generated by the zero modes of $\nop{\psi^a\psi^b}$ and $\nop{\tilde\psi^a\tilde \psi^b}$, $1\le a<b\le 4$.

\bigskip

The fields $V_\lambda(z,\bar z)$ are the vertex operators associated with eigenstates $|\lambda\rangle$ of $\alpha^a_0$ and $\tilde \alpha^b_0$, with eigenvalues $\lambda_L^a$ and $\lambda^b_R$, $a,b=1,\ldots,4$. In each model, the possible vectors of eigenvalues $\lambda\equiv (\vec \lambda_L;\vec \lambda_R):=(\lambda^1_L,\ldots,\lambda^4_L;\lambda^1_R,\ldots,\lambda^4_R)$ form an even self-dual lattice $\Gamma^{4,4}_{\textrm{w--m}}\equiv \Gamma^{4,4}$ (winding-momentum or Narain lattice) with signature $(4,4)$ and quadratic form \beq\label{lattmetric} \lambda\bullet \lambda:= \vec \lambda_L^2-\vec \lambda_R^2=\sum_{a=1}^4 \bigl((\lambda^a_L)^2-(\lambda^a_R)^2)\bigr)\ .\ee 
Explicitly, for each $\lambda\equiv (\vec \lambda_L;\vec \lambda_R)\in \Gamma^{4,4}_{\textrm{w--m}}$, the field $V_\lambda$ is defined as
\begin{equation}\label{vdef}
V_{(\vec\lambda_L;\vec\lambda_R)}(z,{\qu z})
:=\nop{ \exp\left[ i\sum_{a=1}^4\lambda^a_L \phi^a(z) 
+ i\sum_{a=1}^4\lambda^a_R \tilde \phi^a(\qu z)\right] }\; \sigma_{(\vec\lambda_L;\vec\lambda_R)}
\end{equation} where the operators $\sigma_{\lambda}$, $\lambda\in\Gamma^{4,4}_{\textrm{w--m}}$ obey
$$ 
\sigma_\lambda \sigma_\mu = \epsilon(\lambda,\mu)\sigma_{\lambda+\mu}\quad \text{ for all } \lambda,\,\mu\in\Gamma^{4,4}_{\textrm{w--m}}
$$ 
for a suitable normalized $2$-cocycle $\epsilon(\lambda,\mu)\in\{\pm 1\}$ satisfying
\beq\label{local}
\begin{array}{rcl}
\epsilon(\lambda,\mu)&=& (-1)^{\lambda\bullet\mu} \, \epsilon(\mu,\lambda)\ ,\\[0.5em]
\epsilon(\lambda,\mu)\epsilon(\lambda+\mu,\nu) &=& \epsilon(\lambda, \mu+\nu)\epsilon(\mu,\nu)\ ,\\[0.5 em]
\epsilon(\lambda,0)&=&\epsilon(0,\mu) =1\ .
\end{array}
\ee These conditions determine the $2$-cocycle $\epsilon$ up to a $2$-coboundary $\frac{\xi(\lambda)\xi(\mu)}{\xi(\lambda+\mu)}$, for any $\xi:\Gamma^{4,4}_{\textrm{w--m}}\to \{\pm 1\}$ with $\xi(0)=1$, which can be re-adsorbed in the definition of $V_\lambda(z,\bar z)$.
 The OPE of the fields $V_\lambda(z,\bar z)$ with the fermions is non-singular, while
\begin{align}\label{bosOPE1}
j^a(z) V_{\lambda}(w, {\qu w})\  \sim \ 
& \frac{\lambda^a_L}{ z-w} V_{\lambda}(w,{\qu w})\ , \qquad a=1,\ldots, 4, 
\\
\tilde\jmath^a(\qu z) V_{\lambda}(w,{\qu w}) \ \sim \ 
&\frac{ \lambda^a_R}{ \qu z-\qu w} V_{\lambda}(w,{\qu w})\ ,  \qquad a=1,\ldots, 4,
\\
V_{\lambda}(z, {\qu z} ) V_{\lambda^\prime}(w, {\qu w}) \ \sim \ 
&
\epsilon(\lambda,\lambda^\prime)\,
(z-w)^{\vec\lambda_L\cdot\vec\lambda_L^\prime}(\qu z-\qu w)^{\vec\lambda_R\cdot\vec\lambda_R^\prime}
\bigl (V_{\lambda+\lambda^\prime}(w, {\qu w})+\ldots\bigr)\ ,
\label{bosOPE3}\end{align} where $\lambda,\lambda'\in \Gamma^{4,4}_{\textrm{w--m}}$ and $\ldots$ in the last equation denotes less singular terms in $(z-w)$ and $(\qu z-\qu w)$.

\medskip

The real vector space space $\Pi=\Gamma^{4,4}_{\textrm{w--m}}\otimes\RR\cong \RR^{4,4}$ splits into an orthogonal sum $\Pi=\Pi_L\oplus\Pi_R$ of a positive-definite subspace $\Pi_L$, spanned by vectors of the form $(\vec v_L,0)$, and a negative-definite $\Pi_R$ spanned by $(0,\vec v_R)$. Each torus model is uniquely determined by the relative position of $\Pi_L$ and $\Pi_R$ with respect to the lattice $\Gamma^{4,4}_{\textrm{w--m}}$ and we obtain the usual Narain moduli space
$$ \M_{T^4}= O(4,\RR)\times O(4,\RR)\backslash O(4,4,\RR)/O(\Gamma^{4,4})\ .
$$

A concrete realisation of this CFT is given in terms of a supersymmetric non-linear sigma model on a torus $T^4=\RR^4/L$ with a constant antisymmetric B-field $B$. We fix once and for all a set of orthonormal coordinates for the euclidean space $\RR^4$, so that the geometry of the torus is encoded in the shape of the lattice $L\subset \RR^4$. The four scalar fields $X^a(z,\qu z)=\phi^a(z)+\tilde\phi^a(\qu z)$, $a=1,\ldots,4$, describing the target space coordinates, are related to the CFT currents by $j^a=i\partial X^a$, $\tilde j^a=i\bar\partial X^a$. The corresponding 
lattice of $(\alpha_0,\tilde\alpha_0)$-eigenvalues is given by
$$\left\{(\vec\lambda_L,\vec\lambda_R):=\tfrac{1}{\sqrt{2}}(\vec m-B\vec l-\vec l,\vec m- B\vec l+\vec l)\;\mid\; \vec l\in L ,\ \vec m\in L^*\right\}\cong \Gamma^{4,4}_{\textrm{w--m}}\ ,
$$ where $L^*\subset \RR^4$ denotes the dual lattice (here and in the following, the space $\RR^4= L\otimes\RR$ and its dual $(\RR^4)^*= L^*\otimes\RR$ are identified through the Euclidean metric).
The fermionic fields $\psi^a,\tilde\psi^a$ are, up to normalization, the supersymmetric partners of the scalars $X^a$. 

\bigskip

Each of the abstract conformal field theories, defined in terms of the relative position of $\Pi_L$ and $\Pi_R$ with respect to the lattice $\Gamma^{4,4}_{\textrm{w--m}}$, is reproduced by infinitely many different geometric descriptions in terms of non-linear sigma models.  Technically, the choice of any such description amounts to choosing two sublattices  $\Lambda_\textrm{m}$ and $\Lambda_\textrm{w}$ of $\Gamma^{4,4}_{\textrm{w--m}}$, such that $\Lambda_\textrm{m}^\perp=\Lambda_\textrm{m}$, $\Lambda_\textrm{w}^\perp=\Lambda_\textrm{w}$ (i.e., they are maximal isotropic sublattices) and $\Lambda_\textrm{m}\oplus\Lambda_\textrm{w}\cong\Gamma^{4,4}_{\textrm{w--m}}$ \cite{nawe01}. Different choices are related to one another by automorphisms of $\Gamma^{4,4}_{\textrm{w--m}}$ (interpreted as T-dualities). 

In particular, the choice of $\Lambda_\textrm{m}$ determines an isometry $\gamma:\Pi_L\to\Pi_R(-1)$,  characterised by the property that $\gamma(\vec\lambda_L)=\vec\lambda_R$  for each $\lambda=(\vec\lambda_L,\vec\lambda_R)\in\Lambda_\textrm{m}\subset \Pi_L\oplus\Pi_R$ (here, $\Pi_R(-1)$ is obtained from $\Pi_R$ by flipping the sign of the metric). Then one can define a positive definite  lattice $L^*\subset \Pi_L\cong\Pi_R(-1)\cong \RR^4$ of rank four by
\beq\label{momenta} L^*:=\left\{\vec{l}^*\in \RR^4\mid \frac{1}{\sqrt{2}}(\vec{l}^*,\vec{l}^*)\in\Lambda_\textrm{m}\right\}\ .\ee 
The dual lattice $L=(L^*)^*\subset  \RR^4$ is interpreted as the lattice of winding vectors, so that the CFT can be described as a non-linear sigma model on the torus $T^4:= \RR^4/L$.\footnote{From a different viewpoint,  given the interpretation of a left-moving current $j^a$ as a derivative $i\partial X^a$ of a scalar field,  the isometry $\gamma:\Pi_L\to\Pi_R(-1)$ determines which right-moving current $\tilde \jmath^a=\gamma(j^a)$ should correspond to $i\bar\partial X^a$.} 

Finally, the choice of a constant background B-field is encoded in the choice of the lattice $ \Lambda_\textrm{w}$. Indeed, for a given $\Lambda_\textrm{m}$, the most general isotropic sublattice $\Lambda_\textrm{w}\subset \Gamma^{4,4}_{\textrm{w--m}}$ satisfying $\Lambda_\textrm{m}\oplus \Lambda_\textrm{w}\cong\Gamma^{4,4}_{\textrm{w--m}}$ is of the form
$$ \Lambda_\textrm{w}=\{\frac{1}{\sqrt 2}(-B\vec l+\vec l,-B\vec l-\vec l)\,\mid\,\vec l\in L\}\ .$$ Here, the linear map $B$ is given by a real antisymmetric matrix $B_{ij}=-B_{ji}$ as
$$ B \vec{l}_i:=\sum_j B_{ij} \vec{l}^{*j}\ ,
$$ in terms of a basis $\vec{l}_1,\ldots,\vec{l}_4$ of $L$ and of the dual basis $\vec{l}^{*1},\ldots,\vec{l}^{*4}$ of $L^*$. The matrix $B_{ij}$, and consequently the lattice $\Lambda_\textrm{w}$, is determined by the decomposition $\Gamma^{4,4}_{\textrm{w--m}}\otimes\RR =\Pi_L\oplus\Pi_R$ and by the choice of $\Lambda_\textrm{m}\cong \Gamma^{4,4}_{\textrm{w--m}}$, up to a shift by an integral antisymmetric matrix.

\subsection{The `small' $\N=4$ superconformal algebra}\label{s:smallN4}

In the previous section, we have mentioned that every supersymmetric non-linear sigma model on $T^4$ contains a `small' $\N=(4,4)$ superconformal algebra.
The holomorphic $\N=4$ algebra is generated by four supercurrents  $G^{\pm}(z),{G'}^{\pm}(z)$ of weight $(3/2,0)$, the stress energy tensor $T(z)$ and a $su(2)_1$ Kac-Moody algebra (`R-symmetry') whose currents $J^3(z),J^+(z),J^-(z)$ are contained in the `fermionic' $so(4)_1$. It is useful to introduce the complex fields
\begin{align}\label{zdef}
\partial Z^{(1)}(z):=&\frac{1}{\sqrt 2} (j^1(z)+ij^3(z))\ , &
\partial Z^{(1)\,\ast}(z):=&\frac{1}{\sqrt 2} (j^1(z)-ij^3(z))\ , \\
\partial Z^{(2)}(z):=&\frac{1}{\sqrt 2} (j^2(z)+ij^4(z)) \ ,&
\partial Z^{(2)\,\ast}(z):=&\frac{1}{\sqrt 2} (j^2(z)-ij^4(z) \ ,
\end{align}
and
\begin{align}
 \label{Diracblock1}
\chi^{1}(z)&:=\frac{1}{\sqrt 2} (\psi^1(z)+i\psi^3(z))\ ,&  
{\chi^*}^{1}(z)&:=\frac{1}{\sqrt 2} (\psi^1(z)-i\psi^3(z))\ ,\\
\chi^{2}(z)&:=\frac{1}{\sqrt 2} (\psi^2(z)+i\psi^4(z))\ ,&  
{\chi^*}^{2}(z)&:=\frac{1}{\sqrt 2} (\psi^2(z)-i\psi^4(z))\ .
\end{align} Then, the holomorphic $\N=4$ supercurrents are given by
\begin{align}\label{Gpl}
G^{+} (z)& =   \sqrt{2}i \,  (
 \nop{ {\chi^*}^{1} (z)\partial Z^{(1)}(z)}+
 \nop{ {\chi^*}^{2}(z)\partial Z^{(2)}(z)}) \ , \vspace*{0.2cm} \\
G^{-} (z)& =   \sqrt{2}i \,  (
 \nop{\chi^{1} (z)\partial  Z^{(1)\,\ast}(z)}+
 \nop{\chi^{2}(z)\partial Z^{(2)\,\ast}(z)})\ ,  \vspace*{0.2cm} \\  
 {G'}^{+} (z)& =   \sqrt{2} \,  (
 -\nop{{\chi^*}^{1} (z)\partial Z^{(2)\,\ast}(z)}+
 \nop{{\chi^*}^{2}(z)\partial Z^{(1)\,\ast}(z)}) \ ,\vspace*{0.2cm} \\
{G'}^{-} (z)& =   \sqrt{2} \,  (
 \nop{\chi^{1} (z) \partial Z^{(2)}(z)}-
 \nop{\chi^{2}(z) \partial Z^{(1)}(z)}) \ , \label{Gmin}
\end{align} and the $su(2)_1$ `R-symmetry' currents by
\begin{align} J^{3}(z)  
& := \frac{1}{2} \Bigl(\nop{{\chi^*}^{1}(z) \chi^{1} (z)} + \nop{ {\chi^*}^{2} (z) \chi^{2}(z)} \Bigr)\label{Rsymm}\\[2pt]
J^{+}(z) &:= i\nop{{\chi^*}^{1}(z) \, {\chi^*}^{2}(z) } \ , \qquad
J^{-}(z) :=  i  \nop{\chi^{1}(z) \, \chi^{2}(z)} \ .\label{Rsymm1}
\end{align} 

We denote by $SU(2)_{J}$ and $SU(2)_{\tilde{J}}$ the R-symmetry groups, i.e. the groups of symmetries of the CFT generated by the zero modes $J_0^3, J_0^{\pm}$ and by their right-moving counterparts $\tilde J_0^3, \tilde J_0^{\pm}$. It is also useful to define the currents
\begin{align} A^{3}(z)  
& := \frac{1}{2} \Bigl(\nop{{\chi^*}^{1}(z) \chi^{1} (z)} - \nop{ {\chi^*}^{2} (z) \chi^{2}(z)} \Bigr)\label{curr}\\[2pt]
A^{+}(z) &:= i\nop{{\chi^*}^{1}(z) \, {\chi}^{2}(z) } \ , \qquad
A^{-}(z) :=  i  \nop{\chi^{1}(z) \, {\chi^*}^{2}(z)} \ .\label{curr1}
\end{align} and their right-moving counterparts $\tilde A_0^3, \tilde A_0^{\pm}$. These currents form a second $su(2)_1$ affine algebra commuting with \eqref{Rsymm}--\eqref{Rsymm1}. The zero modes $A_0^3, A_0^{\pm}$ and $\tilde A_0^3, \tilde A_0^{\pm}$ generate the groups of symmetries $SU(2)_A$ and $SU(2)_{\tilde A}$, respectively, that commute with $SU(2)_{J}$ and $SU(2)_{\tilde J}$. Altogether, these groups generate the whole fermionic symmetries \eqref{so4ferm}
$$ SO(4)_{f,L}=(SU(2)_{J}\times SU(2)_{A})/(-1)^{J_0^3+A_0^3}\ ,\qquad SO(4)_{f,R}=(SU(2)_{\tilde J}\times SU(2)_{\tilde A})/(-1)^{\tilde J_0^3+\tilde A_0^3}\ ,
$$ where we notice that the central elements  $ (-1)^{J_0^3+A_0^3}\in SU(2)_{J}\times SU(2)_{A}$ and $ (-1)^{\tilde J_0^3+\tilde A_0^3}\in SU(2)_{\tilde J}\times SU(2)_{\tilde A}$ act trivially in the NS-NS sector.

\subsection{A first classification of symmetries}

Our goal is to classify all possible groups $G$ of symmetries of the OPE that fix the small $\N=(4,4)$ superconformal algebra. We focus only on symmetries that act non-trivially on the NS-NS sector; the induced action on the R-R sector is discussed in section \ref{s:Ramond}.

Let us consider first the symmetries that act trivially on the fermionic fields $\psi^a,\tilde\psi^a$ as well as on the bosonic currents $j^a,\tilde j^a$, $a=1,\ldots,4$, so that the $\N=(4,4)$ algebra is automatically invariant. These symmetries act non-trivially only on the fields $V_\lambda$ and the only linear transformations preserving the OPE \eqref{bosOPE1}-\eqref{bosOPE3} are
\beq\label{U1transf} V_\lambda(z,\bar z)\mapsto e^{2\pi i (\vec v_L\cdot \vec\lambda_L-\vec v_R\cdot \vec\lambda_R)} V_\lambda(z,\bar z)=e^{2\pi i \sum_a ( v_L^aj_0^a - v_R^a\tilde j_0^a)} V_\lambda(z,\bar z)\ ,
\ee for some $(\vec v_L,\vec v_R)\in (\Gamma^{4,4}\otimes\RR)/\Gamma^{4,4}$. Therefore, the normal subgroup of  $G$ fixing the fields $\psi^a,\tilde\psi^a,j^a,\tilde j^a$, $a=1,\ldots,4$,  is the group  $U(1)^4_L\times U(1)^4_R$ generated by $j^a_0$ and $\tilde j^a_0$ as in \eqref{u1bos}.  Thus, $G$ can always be written as a product  \beq\label{obvious} G=(U(1)^4_L\times U(1)^4_R).G_0\ ,\ee where $G_0:=G/(U(1)^4_L\times U(1)^4_R)$.

\medskip

We  have reduced our problem to the classification of the possible groups $G_0$ that act non-trivially on the fundamental $u(1)$ currents and fermions.  In order to preserve the OPEs \eqref{fundOPEs} and to fix the supercurrent
$$ \frac{1}{\sqrt{2}}(G^+(z)+G^-(z))=\sum_{a=1}^4 \nop{\psi^a(z)j^a(z)}\ ,
$$ and its right-moving analogue, a symmetry $g$ must act by a simultaneous orthogonal transformation on the fermions and on the $u(1)$ currents, i.e.
\begin{align}
\label{bostransf} j^a(z)&\mapsto \sum_b (g_L)_{ab} j^b(z)\ , &\tilde\jmath^a(z)&\mapsto \sum_b (g_R)_{ab} \tilde \jmath^b(z)\ , \\
 \psi^a(z)&\mapsto \sum_b (g_L)_{ab} \psi^b(z)\ ,& \tilde\psi^a(z)&\mapsto \sum_b (g_R)_{ab} \tilde \psi^b(z)\ ,\label{fermiotransf}
\end{align} where $g_L, g_R\in O(4,\RR)$ are real orthogonal matrices. We denote by $O(4)_L\times O(4)_R$ the group of transformations $(g_L,g_R)$. In order for the $j^aV_\lambda$ OPE to be preserved, $g$ must act on the fields $V_\lambda$ by
\beq\label{Vtransf} V_\lambda(z,\qu z) \mapsto \xi_g(\lambda) V_{g^{-1}(\lambda)}(z,\qu z)\ ,\qquad \lambda\in\Gamma^{4,4}\ ,
\ee where 
\beq\label{lambdatransf} g^{-1}(\vec\lambda_L,\vec\lambda_R):= ((g_L)^{-1}\vec\lambda_L,(g_R)^{-1}\vec \lambda_R)\ ,\qquad (\vec\lambda_L,\vec\lambda_R)\in \Gamma^{4,4}\subset \Pi_L\oplus\Pi_R\ .
\ee The function $\xi_g:\Gamma^{4,4}\to \{\pm 1\}$ satisfies
\beq\label{laxi} \epsilon(\lambda,\mu)=\epsilon(g^{-1}(\lambda),g^{-1}(\mu))\frac{\xi_g(\lambda)\xi_g(\mu)}{\xi_g(\lambda+\mu)}\ ,
\ee for all $\lambda,\mu\in \Gamma^{4,4}$. Since both $\epsilon(g^{-1}(\lambda),g^{-1}(\mu))$ and $\epsilon(\lambda,\mu)$ obey \eqref{local}, a function $\xi_g$ satisfying \eqref{laxi} always exists. In turn,  eq.\eqref{laxi} determines $\xi_g$ up to composition with elements of order two in the group $U(1)^4_L\times U(1)^4_R$ of \eqref{obvious}.\footnote{Technically, eq.\eqref{local} determines the cohomology class of the cocycle $\epsilon$ in $H^2(\Gamma^{4,4},\ZZ_2)\cong H^2(\ZZ^8_2,\ZZ_2)\cong \ZZ_2^{\binom{8}{2}}$, where $\Gamma^{4,4}$ is isomorphic to $\ZZ^8$ as an abelian group. Thus, $\epsilon(g(\lambda),g(\mu))$ must differ from $\epsilon(\lambda,\mu)$ by a $2$-coboundary $d\xi_g$. The cochain $\xi_g$ is determined up to a $1$-cocycle, i.e. a homomorphism $\Gamma^{4,4}\to \ZZ_2^8$.} 

The transformations \eqref{bostransf},\eqref{fermiotransf} and \eqref{Vtransf} define a symmetry of the CFT if and only if the transformation \eqref{lambdatransf} is an automorphism of the lattice $\Gamma^{4,4}$, i.e. an element of $O(\Gamma^{4,4})$. The condition that a certain pair $(g_L,g_R)$ of orthogonal transformations  of the spaces $\Pi_L$ and $\Pi_R$ of left- and right-moving momenta induces an automorphism  of the lattice $\Gamma^{4,4}$ clearly depends on the moduli.

\bigskip

So far we have only required the transformation $g$ to preserve the OPE and fix one holomorphic and one antiholomorphic supercurrent. In order for $g$ to fix the whole $\N=(4,4)$ superconformal algebra, a necessary condition is that the currents \eqref{Rsymm}-\eqref{Rsymm1} be invariant under $g$. 
Notice that if $\det g_L=-1$, then the orthogonal transformation \eqref{fermiotransf} exchanges the two $su(2)_1$ algebras \eqref{Rsymm}--\eqref{Rsymm1} and \eqref{curr}--\eqref{curr1}; a similar exchange occurs for the right-moving currents when $\det g_R=-1$. Therefore, it is sufficient to consider symmetries \eqref{fermiotransf}--\eqref{lambdatransf} within $SO(4)_{L}\times SO(4)_{R}$. With this restriction, it is easy to see that the group of transformations \eqref{fermiotransf} fixing the currents \eqref{Rsymm}-\eqref{Rsymm1} is the $SU(2)_A\times SU(2)_{\tilde A}$ group generated by the zero modes $A_0^3$, $A_0^\pm$ of the currents \eqref{curr}--\eqref{curr1} and their right-moving analogues $\tilde A_0^3$, $\tilde A_0^\pm$. 

We conclude that a transformation $(g_L,g_R)\in O(4)_L\times O(4)_R$, acting on the fields as in \eqref{bostransf}, \eqref{fermiotransf} and \eqref{Vtransf}, fixes the $\N=(4,4)$ superconformal algebra if and only if it belongs to a $SU(2)_L\times SU(2)_R$ subgroup of $O(4)_L\times O(4)_R$, whose action on the complex fields \eqref{zdef} and \eqref{Diracblock1} is
\begin{align}\label{first}
\begin{pmatrix}
\chi^1\\ \chi^2
\end{pmatrix}&\mapsto \rho(g_L)\begin{pmatrix}
\chi^1\\ \chi^2
\end{pmatrix}\ ,& \begin{pmatrix}
{\chi^*}^1\\ {\chi^*}^2
\end{pmatrix}&\mapsto \rho(g_L)^*\begin{pmatrix}
{\chi^*}^1\\ {\chi^*}^2
\end{pmatrix}\ ,\\
\begin{pmatrix}
\partial Z^{(1)}\\ \partial Z^{(2)}
\end{pmatrix}&\mapsto \rho(g_L)\begin{pmatrix}
\partial Z^{(1)}\\ \partial Z^{(2)}
\end{pmatrix}\ ,& \begin{pmatrix}
\partial Z^{(1)\ast}\\ \partial Z^{(2)\ast}
\end{pmatrix}&\mapsto \rho(g_L)^*\begin{pmatrix}
\partial Z^{(1)\ast}\\ \partial Z^{(2)\ast}
\end{pmatrix}\ ,\label{second}\\
\begin{pmatrix}
\tilde \chi^1\\ \tilde \chi^2
\end{pmatrix}&\mapsto \tilde \rho(g_R)\begin{pmatrix}
\tilde \chi^1\\ \tilde \chi^2
\end{pmatrix}\ ,& 
\begin{pmatrix}
\tilde{\chi}^{*1}\\ 
\tilde{\chi}^{*2}
\end{pmatrix}&\mapsto 
\tilde \rho(g_R)^*\begin{pmatrix}
\tilde{\chi}^{*1}\\ \tilde{\chi}^{*2}
\end{pmatrix}\ ,\label{third}
\\
\begin{pmatrix}
\bar\partial Z^{(1)}\\ \bar\partial Z^{(2)}
\end{pmatrix}&\mapsto \tilde \rho(g_R)\begin{pmatrix}
\bar\partial Z^{(1)}\\ \bar\partial Z^{(2)}
\end{pmatrix}\ ,& \begin{pmatrix}
\bar\partial Z^{(1)\ast}\\ \bar\partial Z^{(2)\ast}
\end{pmatrix}&\mapsto \tilde \rho(g_R)^*\begin{pmatrix}
\bar\partial Z^{(1)\ast}\\ \bar\partial Z^{(2)\ast}\\
\end{pmatrix}\ .\label{last}
\end{align}
Here, $\rho(g_L)$ and $\tilde\rho(g_R)$ are $SU(2)$ matrices and $\rho(g_L)^*$ and $\tilde\rho(g_R)^*$ are their complex conjugate. It is straightforward to verify that these transformations leave all generators of the $\N=(4,4)$ superconformal algebra invariant. 

\medskip

It is also useful to give a description of the group $SU(2)_L\times SU(2)_R$ in terms of quaternions. Let $\ii$, $\jj$, $\kk$ be imaginary units satisfying the usual quaternionic multiplication rule
$$ \ii\jj=-\jj\ii=\kk\ ,\quad \jj\kk=-\kk\jj=\ii\ ,\quad \kk\ii=-\ii\kk=\jj\ ,\quad \ii^2=\jj^2=\kk^2=-1\ .
$$ We have the standard identification of the space of quaternions $\HH$ with $\RR^4$
$$ \HH=\{a_1+a_2\ii+a_3\jj+a_4\kk\mid (a_1,\ldots,a_4)\in \RR^4\}\cong \RR^4\ ,
$$ so that we can think of $\Gamma^{4,4}$ as a lattice within $\HH\oplus\HH$ by
$$ (\vec\lambda_L;\vec\lambda_R)\equiv \bigl(\lambda^1_L+\lambda^2_L\,\ii+\lambda^3_L\,\jj+\lambda^4_L\,\kk\; ;\;\lambda_R^1+\lambda^2_R\,\ii+\lambda^3_R\,\jj+\lambda_R^4\,\kk\bigr)\ .
$$
Furthermore, the group unit quaternions forms a copy of $SU(2)$
$$  SU(2)=\{\left(\begin{smallmatrix}
a+id & b+ic\\ -b+ic & a-id
\end{smallmatrix}
\right)\mid a^2+b^2+c^2+d^2=1\}=\{a+b\,\ii+c\,\jj+d\,\kk\in \HH\mid a^2+b^2+c^2+d^2=1\}\ .
$$ Under this identification, the pair $(g_L,g_R)\in SU(2)_L\times SU(2)_R$ can be regarded as a pair of unit quaternions and the action on the fields is simply given by (left) multiplication 
\begin{align}
\label{prima}\psi^1+\psi^2\ii+\psi^3\jj+\psi^4\kk\quad&\mapsto\quad g_L(\psi^1+\psi^2\ii+\psi^3\jj+\psi^4\kk)\ ,\\[3pt]
\tilde\psi^1+\tilde\psi^2\ii+\tilde\psi^3\jj+\tilde\psi^4\kk\quad&\mapsto\quad g_R(\tilde\psi^1+\tilde\psi^2\ii+\tilde\psi^3\jj+\tilde\psi^4\kk)\ ,\\[3pt]
j^1+j^2\ii+j^3\jj+j^4\kk\quad&\mapsto \quad g_L(j^1+j^2\ii+j^3\jj+j^4\kk)\ ,\\[3pt]
\tilde j^1+\tilde j^2\ii+\tilde j^3\jj+\tilde j^4\kk\quad&\mapsto \quad g_R(\tilde j^1+\tilde j^2\ii+\tilde j^3\jj+\tilde j^4\kk)\ ,\\[3pt]
V_\lambda \quad&\mapsto \quad\xi_g(\lambda) V_{g^{-1}(\lambda)}\ ,\label{ultima}
\end{align}
where
\beq
g(\vec\lambda_L,\vec\lambda_R):= (g_L\vec\lambda_L,g_R\vec\lambda_R)\ ,\qquad (\vec\lambda_L,\vec\lambda_R)\in \Gamma^{4,4}\subset \HH\oplus\HH\ .\label{mezzo}
\ee
Once again, eqs.\eqref{prima}--\eqref{ultima} (or, equivalently, \eqref{first}--\eqref{last}) define a symmetry of the CFT if and only if the induced transformation \eqref{mezzo} is an automorphism of the lattice $\Gamma^{4,4}$. This condition constrains the group $G_0$ to be a discrete subgroup of $SU(2)_L\times SU(2)_R$. We stress that the action \eqref{ultima} of $(g_L,g_R)\in G_0$ on the fields $V_\lambda$ is only defined modulo $U(1)^4_L\times U(1)^4_R$ transformations.

\medskip

To summarize, we have found the following characterization of the group $G$:

\begin{theorem}
The group $G$ of symmetries of a supersymmetric non-linear sigma model on $T^4$ preserving a small $\N=(4,4)$ superconformal algebra generated by \eqref{Gpl}--\eqref{Rsymm1}, is given by a product
$$ G=(U(1)^4_L\times U(1)^4_R).G_0\ .
$$ The normal subgroup $U(1)^4_L\times U(1)^4_R$ is generated by the zero modes of the bosonic currents $j^a,\tilde j^a$, $a=1,\ldots,4$, and the quotient $G_0 =G/(U(1)^4_L\times U(1)^4_R)$ is the finite group of elements $(g_L,g_R)\in SU(2)_L\times SU(2)_R$ such that \eqref{mezzo} is an automorphism of the winding-momentum lattice $\Gamma^{4,4}$.   In particular,  $(g_L,g_R)\in G_0$ acts as in \eqref{prima}--\eqref{ultima} on the fields.
\end{theorem} 

In other words, $G_0$ is the intersection of the compact group $SU(2)_L\times SU(2)_R$ and the discrete group $SO(\Gamma^{4,4})$, when both groups are embedded in the orthogonal group $SO(4,4,\RR)$ acting on $\Pi_L\oplus\Pi_R\cong \RR^{4,4}$. Clearly, these embeddings and therefore the intersection $G_0$ depend on the moduli, i.e. on the relative position of the subspaces $\Pi_L$ and $\Pi_R$ with respect to the lattice $\Gamma^{4,4}$.

\bigskip

The group $G_0$ always contains a central $\ZZ_2$ subgroup generated by the involution $(-1,-1)\in SU(2)_L\times SU(2)_R$ that flips the signs of all bosonic currents and all fermions; this is indeed a symmetry of all torus models. On the other hand, a generic deformation of the model will break any symmetry $g\in G_0$ not contained in this central $\ZZ_2$ subgroup. We conclude that the generic symmetry group $G$ is
$$ G=(U(1)^4_L\times U(1)^4_R):\ZZ_2\ .
$$
In the following sections, we will give a complete classification of the possible discrete subgroups $G_0\subset SU(2)_L\times SU(2)_R$ and of the corresponding torus models.

\section{Ramond-Ramond sector, D-branes and triality}\label{s:RRDb}

In this section, we consider the action of the symmetry group $G$ on the Ramond-Ramond fields and on the lattice of D-brane charges. This will lead to a useful characterization of the possible symmetry groups $G$ in section \ref{s:symmetries}.

\subsection{Representation of the symmetries on the Ramond-Ramond sector}\label{s:Ramond}

The Ramond-Ramond sector of the sigma model forms a representation of the algebra of fermionic zero modes $\psi^a_0,\tilde\psi^a_0$. In particular, the ground states have conformal weight $(\frac{1}{4},\frac{1}{4})$ and span a $16$ dimensional space $\Hh$. A convenient basis is given by simultaneous eigenvectors $$ |s_1,s_2;\tilde s_1,\tilde s_2\rangle\ ,\qquad\qquad s_1,s_2,\tilde s_1,\tilde s_2\in\{\pm\tfrac12\}\ ,
$$ of the generators $J_0^3, A_0^3,\tilde J_0^3,\tilde A_0^3$ of the `fermionic' $su(2)_1$ algebras \eqref{Rsymm}, \eqref{Rsymm1}, \eqref{curr}, \eqref{curr1}, such that
\begin{align*}
J_0^3 |s_1,s_2;\tilde s_1,\tilde s_2\rangle&=(s_1+s_2)|s_1,s_2;\tilde s_1,\tilde s_2\rangle\ , &
A_0^3 |s_1,s_2;\tilde s_1,\tilde s_2\rangle&=(s_1-s_2)|s_1,s_2;\tilde s_1,\tilde s_2\rangle\ ,\\[5pt]
\tilde J_0^3 |s_1,s_2;\tilde s_1,\tilde s_2\rangle&=(\tilde s_1+\tilde s_2)|s_1,s_2;\tilde s_1,\tilde s_2\rangle\ , &
\tilde A_0^3 |s_1,s_2;\tilde s_1,\tilde s_2\rangle&=(\tilde s_1-\tilde s_2)|s_1,s_2;\tilde s_1,\tilde s_2\rangle\ .\end{align*}
The left- and right-moving world-sheet fermion number operators are defined by 
\begin{align} &(-1)^{F_L}:=(-1)^{J_0^3}=(-1)^{s_1+s_2}=4\,\psi^1_0\psi^2_0\psi^3_0\psi^4_0\ ,\\ &(-1)^{F_R}:=(-1)^{\tilde J_0^3}=(-1)^{\tilde s_1+\tilde s_2}=4\,\tilde\psi^1_0\tilde\psi^2_0\tilde\psi^3_0\tilde\psi^4_0\ ,\end{align}  and are preserved by the `fermionic' $SO(4)_{f,L}\times SO(4)_{f,R}$ transformations. We denote by $\Pi_\textrm{even}$ and $\Pi_\textrm{odd}$ the subspaces of R-R ground states with $(-1)^{F_L+F_R}=1$ and $(-1)^{F_L+F_R}=-1$, respectively.

\medskip

Let us consider the action of the group $G$ on the R-R ground states. Clearly, the normal subgroup $U(1)^4_L\times U(1)^4_R$ acts trivially on this space, so we can focus on the quotient $G_0$.
An element $g=(g_L,g_R)\in G_0$ acts by an $ SU(2)_A\times SU(2)_{\tilde A}$ transformation on the fermionic fields $\psi^a,\tilde\psi^a$ and the action is represented by some $SU(2)$ matrices $\rho(g_L)$, $\tilde\rho(g_R)$ as in \eqref{first}--\eqref{last}. In order to preserve the OPE, $g$ has to act by the same 
$ SU(2)_A\times SU(2)_{\tilde A}$ transformation also on the R-R ground states.
Assume, for the sake of simplicity, that this transformation is contained in the Cartan  torus generated by $A_0^3$ and $\tilde A_0^3$. If the eigenvalues of $\rho(g_L)$ and $\tilde\rho(g_R)$ are $\zeta_L,\zeta_L^{-1},\zeta_R,\zeta_R^{-1}$, then the action of $g$ in the R-R sector is given by
$$ g|s_1,s_2;\tilde s_1,\tilde s_2\rangle=\zeta_L^{A_0^3}\zeta_R^{\tilde{ A}_0^3}|s_1,s_2;\tilde s_1,\tilde s_2\rangle=\zeta_L^{s_1-s_2}\zeta_R^{\tilde s_1-\tilde s_2}|s_1,s_2;\tilde s_1,\tilde s_2\rangle\ .
$$
\begin{table}[t]$$ \begin{array}{c|c|c}
(-1)^{F_L} & (-1)^{F_R} & \text{Eigenvalues of }g\\
\hline
-1 & -1 & 1\ ,\quad 1\ ,\quad 1\ ,\quad 1\ ,\\
+1 & +1 & \zeta_L\zeta_R\ ,\quad \zeta_L^{-1}\zeta_R\ ,\quad \zeta_L\zeta_R^{-1}\ ,\quad \zeta_L^{-1}\zeta_R^{-1}\\
+1 & -1 & \zeta_L\ ,\quad \zeta_L\ ,\quad \zeta_L^{-1}\ ,\quad\zeta_L^{-1}\ ,\\
-1 & +1 & \zeta_R\ ,\quad\zeta_R\ ,\quad\zeta_R^{-1}\ ,\quad \zeta_R^{-1}
\end{array}
$$\caption{\small The eigenvalues of $g$ on the eigenspaces of $(-1)^{F_L}$ and $(-1)^{F_R}$.}\label{t:eigeng}\end{table}
Hence, the eigenvalues of $g$ on the eigenspaces of $(−1)^{F_L}$ and $(−1)^{F_R}$ are as in table \ref{t:eigeng}.
Notice that the eigenspace with $(-1)^{F_L}=(-1)^{F_R}=-1$ is fixed pointwise by $g$. Furthermore, the central involution $(-1,-1)\in SU(2)_L\times SU(2)_R$ acts trivially on the whole space $\Pi_\textrm{even}$ with positive total fermion number $(-1)^{F_L+F_R}=+1$. Thus, the group acting faithfully on $\Pi_\textrm{even}$ is $G_1=G_0/(-1,-1)$ which is a subgroup of $(SU(2)_L\times SU(2)_{R})/(-1,-1)$. These conclusions are valid even when the action of $g$ on the fermions is not generated by $A_0^3$ and $\tilde A_0^3$,  since by a suitable conjugation within $SU(2)_A\times SU(2)_{\tilde A}$ one can always reduce $g$ to this Cartan torus while preserving the eigenspaces of $(-1)^{F_L}$ and $(-1)^{F_R}$. As will be described in the following sections, it is easier to characterize the possible groups $G$ of symmetries by considering their action on the eigenspace $\Pi_\textrm{even}$ with $(-1)^{F_L+F_R}=+1$. 

\medskip

In the above derivation we have been slightly naive in the identification of the action of $g$ on the R-R sector. In general, a $SO(4)_{f,L}\times SO(4)_{f,R}$ transformation on the fundamental fermions in the NS-NS sector determines the transformation on the R-R sector only up to a sign -- the symmetries $(-1)^{J_0^3+A_0^3}$ and $(-1)^{\tilde J_0^3+\tilde A_0^3}$ act trivially on the NS-NS sector and by a minus sign on the R-R sector. Therefore, for a generic subgroup $G$ of $SO(4)_{f,L}\times SO(4)_{f,R}$ acting on the NS-NS sector, the induced group acting on the R-R sector and preserving the OPE is a $\ZZ_2$-central extension of $G$. 

However, the groups $G_0$ we are considering are actually contained in the  subgroup $SU(2)_A\times SU(2)_{\tilde A}$ of $SO(4)_{f,L}\times SO(4)_{f,R}$ generated by $A_0^3,A_0^\pm,\tilde A_0^3,\tilde A_0^\pm$, so that their central extension in the R-R sector is always the trivial one $\ZZ_2\times G_0$. Therefore, we do not lose any information if we simply focus on the component of $\ZZ_2\times G_0$ that fixes the four R-R states with $(-1)^{F_L}=(-1)^{F_R}=-1$. The fields associated with these four states are the generators of a spectral flow isomorphism relating the NS-NS and the R-R sectors. In the context of compactification of type II superstrings, the presence of these fields in the spectrum of the internal CFT is related to unbroken space-time supersymmetries. In particular, the requirement that a symmetry $g$ commutes with space-time supersymmetry implies that these spectral flow generators be fixed by $g$. An analogous condition was considered in the classification of the groups of symmetries of K3 models in \cite{K3symm}.

\subsection{D-branes and triality}\label{s:Dbtrial}

Fundamental D-branes are defined, in a given geometric interpretation of the torus model, by imposing either Neumann or Dirichlet boundary conditions in each direction. Each of these boundary conditions preserve a particular bosonic $u(1)^4$ subalgebra of the original $u(1)^4_L\oplus u(1)^4_R$.  For example, consider a geometric realization of the model, as described in section \ref{s:generalities}, associated with a maximal isotropic sublattice $\Lambda_\textrm{m}\equiv \spa_\ZZ\{\lambda_1,\ldots,\lambda_4\}\subset\Gamma^{4,4}$, which determines the lattice $L^*$ of space-time momenta on a torus $\RR^4/L$ as in eq.\eqref{momenta}. Then, the Dirichlet boundary conditions describing a D0-brane in the Ramond-Ramond sector are enforced by
\begin{align} 
\label{Dbranebos} (\vec\lambda_{iL}\cdot \alpha_n - \vec\lambda_{iR}\cdot \tilde\alpha_{-n})|\mu;\Lambda_\textrm{m},\eta\rangle\!\rangle &=0\ ,\qquad n\in\ZZ,\ i=1,\ldots,4\ ,\\
\label{Dbraneferm} (\vec\lambda_{iL}\cdot \psi_r +i\eta \vec\lambda_{iR}\cdot \tilde\psi_{-r})|\mu;\Lambda_\textrm{m},\eta\rangle\!\rangle &=0\ ,\qquad r\in\ZZ,\ i=1,\ldots,4\ .
\end{align} Here, $\eta\in \{\pm 1\}$ and $\Lambda_\textrm{m}$ determine which $\N=4$ and which $u(1)^4$ subalgebras are preserved by the boundary conditions, $\mu\equiv \frac{1}{\sqrt{2}}({\vec l}^*,{\vec l}^*)\in\Lambda_\textrm{m}$, with ${\vec l}^*\in L^*$, labels the distinct representations of this preserved $u(1)^4$ subalgebra and $|\mu;\Lambda_\textrm{m},\eta\rrangle$ denotes the Ishibashi state in the corresponding sector. The boundary state $\bbar a;\Lambda_\textrm{m},\eta\rrangle$ is then obtained by a superposition of the Ishibashi states for the different values of $\mu \in \Lambda_\textrm{m}$
$$ \bbar a;\Lambda_\textrm{m},\eta\rrangle = \N \sum_{\mu\in \Lambda_\textrm{m}} e^{2\pi i (a\bullet \mu)} |\mu ; \Lambda_\textrm{m},\eta\rrangle\ .
$$ The modulus $a \in (\Lambda_\textrm{w}\otimes \RR)/\Lambda_\textrm{w}$ represents the position of the D0-brane on the four-dimensional torus and $\N$ is a suitable normalization. The treatment of the NS-NS sector is analogous, the only difference being in the boundary conditions \eqref{Dbraneferm}, where the fermion modes $\psi_r$, $\tilde\psi_{-r}$ are defined for $r\in\frac{1}{2}+\ZZ$. In a full ten dimensional superstring theory, space-time supersymmetric D-branes are obtained by tensoring these boundary states with the analogous D-brane states in the remaining $6$ space-time directions and considering a suitable GSO projected combination of NS-NS and R-R contributions.

\medskip

The T-duality group $O(\Gamma^{4,4})$ acts transitively on the set of maximal isotropic sublattices $\Lambda\subset \Gamma^{4,4}$, and by enforcing the conditions \eqref{Dbranebos} and \eqref{Dbraneferm} for a given such $\Lambda$ one recovers all the other fundamental D-branes with different combinations of Neumann and Dirichlet boundary conditions. For example, if we extend the set of generators of the lattice $\Lambda_\textrm{m}$ above to a basis $\lambda_1,\ldots,\lambda_4,\hat\lambda_1,\ldots,\hat\lambda_4$ of $\Gamma^{4,4}$ such that
\beq\label{stand} \lambda_i\bullet\lambda_j=\hat\lambda_i\bullet\hat\lambda_j=0\ ,\qquad \lambda_i\bullet\hat\lambda_j=\delta_{ij}\ ,
\ee then the isotropic sublattice  generated by $\hat\lambda_1,\ldots,\hat\lambda_4$ is associated with a D4-brane and the lattice generated by $\lambda_1,\lambda_2,\hat\lambda_3,\hat\lambda_4$ is associated with a D2-brane extended in the 3--4 directions. Thus, the set of fundamental D-branes can be described, via \eqref{Dbranebos} and \eqref{Dbraneferm}, as the set of maximal isotropic sublattices of $\Gamma^{4,4}$. This description makes no reference to a choice of a geometric interpretation of the model; the latter just amounts to choosing which element in this set should be considered a D0-brane and which one a D4-brane.

\medskip

The Ramond-Ramond charge of a D-brane corresponds to the ground state component of the boundary state $\bbar a;\Lambda,\eta\rrangle$ and is completely determined by eq.\eqref{Dbraneferm} for $r=0$ up to normalization, which in turn is fixed by the Cardy and factorisation conditions.  Let us focus on the case $\eta=1$ and use a simplified notation $\bbar \Lambda \rrangle\equiv \bbar a;\Lambda,\eta=1 \rrangle$, where we drop the dependence on the modulus $a$. We denote by $\Psi_\Lambda$ the R-R ground state component of $\bbar\Lambda \rrangle$.  For $\eta=+1$, equations \eqref{Dbraneferm} read
$$ c(\lambda)\bbar\Lambda\rrangle=c(\lambda)\Psi_\Lambda=0\ ,\qquad \text{for all }\lambda\in \Lambda\subset\Gamma^{4,4}\ ,
$$  where the operators
\beq\label{Cliff} c(v):=\vec v_L\cdot \psi_0+i\vec v_R\cdot\tilde\psi_0\ ,\qquad v\equiv (\vec v_L,\vec v_R)\in\Gamma^{4,4}\otimes\RR\ ,
\ee generate the real Clifford algebra associated with the space $\Gamma^{4,4}\otimes\RR$, with anticommutation relations
\beq\label{Cliff2} \{c(v),c(w)\}=\vec v_L\cdot \vec w_L-\vec v_R\cdot\vec  w_R=v\bullet w\ .
\ee
 The overlap between two D-branes is given by
 \beq\label{RRmetric} \llangle\Lambda'\bbar (-1)^{F_R+1} q^{\frac{L_0+\tilde L_0}{2}-\frac{c}{24}}\bbar\Lambda \rrangle=\langle\Psi_{\Lambda'}| (-1)^{F_R+1} \Psi_\Lambda \rangle\ ,
 \ee which defines a non-degenerate bilinear form of signature $(8,8)$ on the lattice of D-brane charges. 
Since all $c(v)$ anticommute with the chirality operator
$$ c(v)(-1)^{F_L+F_R}=-(-1)^{F_L+F_R}c(v)\ ,\qquad v\in\Gamma^{4,4}\ ,
$$ $\Psi_\Lambda$ must have definite chirality.  D-brane charges with opposite chiralities are orthogonal with respect to \eqref{RRmetric}, so that the lattice of D-brane charges splits as an orthogonal sum $\Gamma^{4,4}_\textnormal{even}\oplus_\perp \Gamma^{4,4}_\textrm{odd}$ of lattices isomorphic to $\Gamma^{4,4}$. In a geometric description of the model, chirality corresponds to the parity of D-brane dimensionalities and the lattices $\Gamma^{4,4}_\textrm{even}$ and $\Gamma^{4,4}_\textrm{odd}$ are identified with the even and odd integral cohomology groups $H^\textnormal{even}(T^4,\ZZ)$ and $H^\textnormal{odd}(T^4,\ZZ)$, and the bilinear form is the cup product.

Each of the $8$-dimensional vector spaces $$ \Pi_\textnormal{even}=\Gamma^{4,4}_\textnormal{even}\otimes\RR\ ,\qquad\qquad   \Pi_\textrm{odd}=\Gamma^{4,4}_\textrm{odd}\otimes\RR\ ,$$ contains two orthogonal subspaces $\Pi^{\pm}_\textrm{even}$, $\Pi^{\pm}_\textrm{odd}$ with definite signature with respect to \eqref{RRmetric} corresponding to the $(-1)^{F_R+1}$-eigenspaces with signature $\pm1$. From now on, we denote by $\Gamma^{4,4}_\textnormal{w--m}$ the lattice of winding-momenta (eigenvalues of $\alpha_0,\tilde\alpha_0$), to distinguish it from the lattices of D-brane charges, and set $\Pi_\textrm{w--m}:=\Pi=\Gamma^{4,4}_\textnormal{w--m}\otimes \RR$.

\bigskip

From a formal point of view, this construction is completely symmetric among the three lattices $\Gamma^{4,4}_\textnormal{w--m},\Gamma^{4,4}_\textnormal{even}$ and $\Gamma^{4,4}_\textrm{odd}$, in the sense that one can start from any of the three lattices to define the other two. More precisely, let us consider one of the three lattices, denoted by $\Gamma^{4,4}_1$ and define the Clifford algebra \eqref{Cliff2} based on the real vector space $\Pi_1=\Gamma^{4,4}_1\otimes\RR\cong \RR^{4,4}$. This Clifford algebra admits a representation as an algebra of real matrices over a  $16$-dimensional real space $V$. As usual, the action of the group $SO(\Pi_1)\cong SO(4,4)$  on the Clifford algebra generators induces a spinorial representation of $\Spin(4,4)$ on $V$.   More precisely, $V$ is the direct sum $V=\Pi_2\oplus \Pi_3$ of two  irreducible $\Spin(4,4)$ representations of opposite chiralities. Here, $\Pi_2$, $\Pi_3$ are the $8$ dimensional eigenspaces of the involution (chirality operator)
$$ (-1)^{F_L+F_R}=16\,c(e_1)\cdots c(e_8)\ ,
$$ with $e_1,\ldots,e_8$ any oriented basis of $\Pi_1$ with $e_i\bullet e_j=\pm \delta_{ij}$.

The action of  $\Spin(4,4)$ preserves a symmetric bilinear form $(\cdot ,\cdot )$ of signature $(4,4)$ on each of the spaces $\Pi_2$ and $\Pi_3$. In order to define this bilinear form, it is useful to choose a basis $\lambda_1,\ldots,\lambda_4,$ $\hat\lambda_1,\ldots,\hat\lambda_4$ of $\Gamma^{4,4}_1$ with standard Gram matrix \eqref{stand}. Then, there exists a one dimensional space of states $\mu\in V$ satisfying
\beq\label{kerno} c(\lambda_i)\mu=0\ ,\qquad i=1,\ldots,4\ .
\ee We choose a non-zero vector $\mu$ in this space and, as usual, we define a set of generators of $V$ by acting on $\mu$ in all possible ways with the operators $c(\hat \lambda_i)$
\beq\label{labase} \mu_{i_1,\ldots, i_r}:=c(\hat\lambda_{i_1})\cdots c(\hat\lambda_{i_r})\mu\ ,\qquad\qquad 1\le i_1<\ldots<i_r\le 4\ ,
\ee and define the bilinear form $(\cdot ,\cdot )$ by
\beq\label{bilin} (\mu_{i_1,\ldots,i_r},\mu_{j_1,\ldots,j_s})=\begin{cases}0 & \text{if }r+s\neq 4,\\
\epsilon_{i_r,\ldots,i_1,j_1,\ldots,j_s} & \text{if } r+s=4 ,
\end{cases}
\ee
where $\epsilon_{ijkl}$ is the completely antisymmetric tensor with $\epsilon_{1234}=1$. In particular,
\beq\label{norma} (\mu , c(\hat\lambda_1)c(\hat\lambda_2)c(\hat\lambda_3)c(\hat\lambda_4)\mu )= 1\ .
\ee 
Notice that $\mu$ has definite chirality, which depends on the orientation of the basis $\lambda_1,\ldots,\lambda_4$, $\hat\lambda_1,\ldots,\hat\lambda_4$. Up to a possible exchange of $\lambda_1$ and $\hat\lambda_1$, we can assume that $(-1)^{F_L+F_R}\mu=+\mu$. Therefore, all vectors $\mu_{i_1,\ldots, i_r}$ of the basis have definite chirality $(-1)^r$, so that \eqref{bilin} defines a bilinear form of signature $(4,4)$ on both $\Pi_2$ and $\Pi_3$. 

We define the lattice $\Gamma^{4,4}_2\oplus \Gamma^{4,4}_3$ as the $\ZZ$-span of the vectors \eqref{labase}, with $\Gamma^{4,4}_i\otimes\RR =\Pi_i$, $i=2,3$. By \eqref{bilin}, $\Gamma^{4,4}_2$ and $\Gamma^{4,4}_3$ are mutually orthogonal even unimodular lattices of signature $(4,4)$. Given any other basis $\lambda'_1,\ldots,\lambda'_4$, $\hat\lambda'_1,\ldots,\hat\lambda'_4$, satisfying \eqref{stand}, we can define a vector $\mu'$ satisfying \eqref{kerno} and (upon possibly reordering the sets $\lambda'_1,\ldots,\lambda'_4$ and $\hat\lambda'_1,\ldots,\hat\lambda'_4$) \eqref{norma}. This vector $\mu'$ is determined by \eqref{kerno} and \eqref{norma} up to a sign. Furthermore, $\mu'$ depends only on the maximal isotropic sublattice $\Lambda'\subset \Gamma^{4,4}_1$ spanned by $\lambda'_1,\ldots,\lambda'_4$, since eq.\eqref{norma} gives the same normalization for all choices of $\hat\lambda'_1,\ldots,\hat\lambda'_4$ satisfying \eqref{stand}.   One finds that $\mu'$ has integral product with all generators of the basis \eqref{labase}. Therefore, by self-duality, $\mu'$ must be also contained in $\Gamma^{4,4}_2\oplus\Gamma^{4,4}_3$. As a consequence, the definition of the lattices $\Gamma^{4,4}_2$ and $\Gamma^{4,4}_3$  does not depend on the choice of the basis $\lambda_1,\ldots,\lambda_4,\hat\lambda_1,\ldots,\hat\lambda_4$. Notice that also all vectors $\mu_{i_1,\ldots,i_r}$ in \eqref{labase} are associated with some maximal  isotropic sublattice of $\Gamma^{4,4}_1$, generated by a suitable subset of four mutually orthogonal vectors in the basis $\lambda_1,\ldots,\lambda_4,\hat\lambda_1,\ldots,\hat\lambda_4$.  

A generic automorphism $g_1 \in SO^+(\Gamma^{4,4}_1)$ of the lattice $\Gamma^{4,4}_1$ lifts to some $\hat g\in \Spin(4,4)$ transformation on $\Pi_2$ and $\Pi_3$. By \eqref{kerno} and \eqref{norma}, this transformation maps a vector $\mu\in \Gamma^{4,4}_2\oplus\Gamma^{4,4}_3$ associated with a maximal isotropic $\Lambda\subset \Gamma^{4,4}_1$, to a vector $\pm \mu'$ associated with $g(\Lambda)$. Thus, $\hat g$ maps the lattice $\Gamma^{4,4}_2\oplus\Gamma^{4,4}_3$ into itself and since it also preserves the chirality and the bilinear form on $V$, it must act by automorphisms $g_2,g_3$ on both $\Gamma^{4,4}_2$ and $\Gamma^{4,4}_3$. 

The choice of an orthogonal decomposition $\Pi_1=\Pi^+_1\oplus\Pi^-_1$ into a positive and a negative definite  oriented subspace yields  the definition of the involutions
$$ (-1)^{F_L}=4\,c(e_1)\cdots c(e_4)\ ,\qquad \qquad
 (-1)^{F_R}=4\,c(\hat e_1)\cdots c(\hat e_4)\ ,
$$ where $e_1,\ldots,e_4$ and $\hat e_1,\ldots,\hat e_4$ are oriented orthonormal bases of $\Pi_1^+$ and $\Pi_1^-$. In turn, this leads to the splittings $\Pi_2=\Pi_2^+\oplus \Pi_2^-$ and $\Pi_3=\Pi_3^+\oplus \Pi_3^-$ into the eigenspaces of $(-1)^{F_R+1}$.

Finally, this construction leads to the definition of a triality
\begin{align}
\label{triality}\begin{array}{ccccccc}
\Gamma^{4,4}_{3} & \times & \Gamma^{4,4}_1& \times& \Gamma^{4,4}_2 &\qquad \to \qquad\qquad  &\ZZ\ ,\\[5pt]
(\mu_3,&& \mu_1,&& \mu_2)&\qquad \mapsto  \qquad\qquad &(\mu_3,c(\mu_1)\mu_2)\ ,\end{array}
\end{align} i.e. a non-degenerate trilinear form,\footnote{Non-degenerate here means that every non-zero vector in one of the three lattices determines a non-degenerate bilinear form on the other two.} which extends to a triality of real vector spaces $\Pi_3 \times \Pi_1\times \Pi_2 \to \RR$. In particular, each $\mu_1\in \Gamma^{4,4}_1$ defines two linear maps $c(\mu_1)_{23}:\Gamma^{4,4}_2\to (\Gamma^{4,4}_3)^*\cong \Gamma^{4,4}_3$ and $c(\mu_1)_{32}:\Gamma^{4,4}_3\to (\Gamma^{4,4}_2)^*\cong \Gamma^{4,4}_2$, which are isomorphisms when $\mu_1 \bullet\mu_1=\pm 2$. In the same fashion, for any permutation $(i,j,k)$ of $(1,2,3)$,  the triality \eqref{triality} defines for each $\mu_i\in \Gamma^{4,4}_i$ a map $c(\mu_i)_{jk}:\Gamma^{4,4}_j\to (\Gamma^{4,4}_k)^*\cong \Gamma^{4,4}_k$. A straightforward computation shows that the maps $c(\mu_i):=c(\mu_i)_{jk}\oplus c(\mu_i)_{kj}$, for all $\mu_i\in \Gamma^{4,4}_i$, generate an algebra of matrices  on the space $\Pi_j\oplus\Pi_k$ that  forms a representation of the Clifford algebra of $\Gamma^{4,4}_i$.
The triality among vector spaces is preserved by the action of $\hat g\in \Spin(4,4)$
$$ (g_3(\mu_3),g_1(\mu_1),g_2(\mu_2))=(\mu_3,\mu_1,\mu_2)\ ,
$$ where $g_1$, $g_2$ and $g_3$ are the vector and the two spinor representations of $\hat g$, respectively. Therefore, the statement that an automorphism $g_1\in O(\Gamma^{4,4}_1)$ induces automorphisms $g_2\in O(\Gamma^{4,4}_2)$ and $g_3\in O(\Gamma^{4,4}_3)$ can also be `symmetrized': for each permutation $(i,j,k)$  of $(1,2,3)$ and for each $\hat g\in \Spin(4,4)$, $g_j$ and $g_k$ are automorphisms of $\Gamma^{4,4}_j$ and $\Gamma^{4,4}_k$ if and only if $g_i$ is an automorphism of $\Gamma^{4,4}_i$. Furthermore, the description of the moduli space of torus models as parametrizing the relative position of an even unimodular lattice $\Gamma^{4,4}$ with respect to a positive and a negative definite subspaces $\Pi^+,\Pi^-$ with $\Gamma^{4,4}\otimes \RR=\Pi^+\oplus_\perp \Pi^-$ holds for any interpretation of $\Gamma^{4,4}$ as the lattice of winding-momenta, even or odd-dimensional D-branes  \cite{nawe01}.

This `democracy' among the three lattices is characteristic of four-dimensional torus models and is related to the triality automorphism of $\Spin(4,4)$, which permutes the vector and the two irreducible spinor representations.

\section{Groups and models}\label{s:groups}

In this section, we classify all possible groups of symmetries $G$ preserving the $\N=(4,4)$ superconformal algebra and describe the corresponding torus models. The main idea is to focus on the action of $G$ on the lattice $\Gamma^{4,4}_\textrm{even}$ of even dimensional D-brane charges. By the triality described in section \ref{s:Dbtrial}, this analysis is sufficient to reconstruct the action of $G$ on the lattice $\Gamma^{4,4}_\textrm{w--m}$ of winding-momenta and on all the fields of the theory.

\subsection{Symmetries of torus models}\label{s:symmetries}

Let $G$ be the group of symmetries of a torus model that preserve the small $\N=(4,4)$ superconformal algebra. A discussed in section \ref{s:Ramond}, the representation of $G$ over the space $\Pi_\textrm{even}$ of R-R ground states with positive chirality is given by a group $G_1\cong G_0/(-1)$ of orthogonal transformations of $\Pi_\textrm{even}$ that fix the subspace $\Pi_\textrm{even}^+$ with $(-1)^{F_R+1}=+1$. Furthermore, $G_1$ must  act by automorphisms on the lattice $\Gamma^{4,4}_\textrm{even}$ of even D-brane charges.

Conversely, let $G_1\subset SO^+(\Gamma^{4,4}_\textrm{even})\subset SO^+(4,4)$ be a group of automorphisms of the lattice of even D-brane charges that fix the subspace $\Pi^+_\textrm{even}\subset \Pi_\textrm{even}$. Thus, $G_1$ acts faithfully by $SO(4)$ transformations on the space $\Pi_\textrm{even}^-$ of R-R ground fields with $(-1)^{F_R+1}=-1$. The spin cover $\Spin(4)$ of this $SO(4)$ group is the $SU(2)_{A}\times SU(2)_{\tilde A}$ group generated by the zero modes of the currents $A_0^3,A_0^{\pm},\tilde A_0^3,\tilde A_0^{\pm}$. Therefore, the group $G_1$ is the representation on the R-R sector of some group $G_0\subset SU(2)_L\times SU(2)_R$, acting on the NS-NS-fields as in \eqref{first}--\eqref{last}. The group $G_0$ is a isomorphic to a double (spin) cover of $G_1$, i.e. to the preimage of $G_1\subset SO(\Gamma^{4,4}_\textrm{even})\subset SO(4,4)$ under the spin covering $\Spin(4,4)\to SO(4,4)$, and acts on the space $\Pi_\textrm{w--m}$ of winding-momenta in one of the irreducible spin representations. By the triality construction described in the previous section, since $G_1$ acts by automorphisms on the lattice $\Gamma^{4,4}_\textrm{even}$, then $G_0$ must act by automorphisms on the lattice $\Gamma^{4,4}_\textrm{w--m}$ of winding momenta. Therefore, $(U(1)^4_L\times U(1)^4_R).G_0$ is a group of symmetries of the torus model that preserves the $\N=(4,4)$ algebra.

\medskip

Thus, we have found an alternative description of the groups of symmetries $G$ considered in section \ref{s:generalities}. In fact, as will be discussed below, there are two more equivalent characterizations of these groups:

\begin{theorem}
Let $G_1$ be a subgroup of $SO(4,\RR)$ and let $\ZZ_2.G_1\subseteq \Spin(4)$ denote its preimage under the spin covering homomorphism $\Spin(4)\to SO(4)$. The following properties are equivalent:
\begin{enumerate}
\item The group 
$$ G\cong (U(1)^4_L\times U(1)^4_R).(\ZZ_2.G_1)\ ,
$$
is the group of symmetries of a supersymmetric non-linear sigma model on $T^4$ that preserves a small $\N=(4,4)$ superconformal algebra.
\item $G_1$ is a subgroup of $SO^+(\Gamma^{4,4}_\textrm{even})\subset SO^+(4,4,\RR)$ that fixes pointwise a positive-definite subspace $\Pi^+_\textrm{even}\subset \Gamma^{4,4}_\textrm{even}\otimes \RR$ of dimension four.
\item $G_1=SO_0(\Lambda)$, where $\Lambda$ is an even positive-definite lattice of rank at most four and $SO_0(\Lambda)\subseteq SO(\Lambda)$ is  the group of automorphisms of $\Lambda$ that act trivially on the discriminant group $\Lambda^*/\Lambda$ ($\Lambda^*$ denotes the dual of $\Lambda$, see appendix \ref{s:lattices}).
\item $G_1$ is the subgroup of $W^+(E_8)$, the group of even Weyl transformations of the $E_8$ lattice,  fixing a sublattice of rank at least four.
\end{enumerate}
\end{theorem}

The equivalence of (1) and (2) has been discussed above. Let us first show that $(2)\Leftrightarrow (3)$. Let $G_1$ be a subgroup of $SO^+(\Gamma^{4,4}_\textnormal{even})$ that leaves the subspace $\Pi^+_\textrm{even}$ pointwise fixed. Let us denote by $\Gamma^{G_1}\subseteq \Gamma^{4,4}_\textrm{even}$  the sublattice of vectors fixed by $G_1$
$$ \Gamma^{G_1}:=\{v\in \Gamma^{4,4}_\textnormal{even} \mid g(v)=v \text{ for all } g\in G_1\}\ ,
$$ and by $\Gamma_{G_1}$ its orthogonal complement in $\Gamma^{4,4}_\textrm{even}$
$$ \Gamma_{G_1}:=\{w\in \Gamma^{4,4}_\textnormal{even} \mid w\bullet v=0 \text{ for all } v\in \Gamma^{G_1}\}\ .
$$ Since $\Pi^+_\textrm{even}\subseteq \Gamma^{G_1}\otimes\RR$, it follows that $\Gamma_{G_1}$ is an 
 even negative-definite lattice of rank at most $4$. Notice that $\Gamma^{G_1}$ and $\Gamma_{G_1}$ are primitive mutually orthogonal sub-lattices of $\Gamma^{4,4}_\textnormal{even}$ and the direct sum $\Gamma^{G_1}\oplus \Gamma_{G_1}$ has maximal rank $8$. Then, by the standard `gluing' construction (see appendix \ref{s:lattices} for details), there is an isomorphism $(\Gamma^{G_1})^*/\Gamma^{G_1}\stackrel{\cong}{\to} (\Gamma_{G_1})^*/\Gamma_{G_1}$ of discriminant groups that reverses the induced discriminant form and such that
 $$ \Gamma^{4,4}_\textnormal{even}=\{(v,w)\in (\Gamma^{G_1})^*\oplus (\Gamma_{G_1})^*\mid [v] \cong [w] \}\ ,
 $$ where for each $x$ in $(\Gamma^{G_1})^*$ or $(\Gamma_{G_1})^*$, $[x]$ denotes the image of $x$ in the corresponding discriminant group. Since $G_1$ acts trivially on the discriminant group $(\Gamma^{G_1})^*/\Gamma^{G_1}$, then it must act trivially also on $(\Gamma_{G_1})^*/\Gamma_{G_1}$ \cite{Nikulin}. We conclude that $G_1$ is a group of automorphisms of the positive-definite even lattice $\Lambda\cong \Gamma_{G_1}(-1)$ of rank (at most) four that fixes the discriminant group $\Lambda^*/\Lambda$.
 
 Vice-versa, given any even positive-definite  lattice $\Lambda$ of rank at most four, there is a primitive embedding of $\Lambda(-1)$ in $\Gamma^{4,4}_\textrm{even}$ (see Theorem 1.12.2 of \cite{Nikulin}) and one can always find a positive definite four dimensional subspace $\Pi^+_\textrm{even}\subset \Gamma^{4,4}_\textnormal{even}\otimes \RR$ such that $\Lambda(-1)= \Gamma^{4,4}_\textnormal{even}\cap (\Pi^+_\textrm{even})^\perp$. As explained in appendix \ref{s:lattices}, the automorphisms of $\Lambda$ that fix the discriminant group $\Lambda^*/\Lambda$ extend to automorphisms of $\Gamma^{4,4}_\textnormal{even}$ that fix pointwise the orthogonal space $\Pi^+_\textrm{even}$. This proves the equivalence of $(2)$ and $(3)$.
 
 \medskip
 
The proof that $(3)\Leftrightarrow (4)$ is completely analogous, with the lattice $\Gamma^{4,4}$ replaced by $E_8$. Let $\Lambda$ be an even positive definite lattice of rank at most $4$ and let $SO_0(\Lambda)$ denotes the group of (positive determinant) automorphisms of $\Lambda$ that act trivially on the discriminant group $\Lambda^*/\Lambda$.
By a theorem by Nikulin (Theorem 1.12.4 of \cite{Nikulin}), any even positive definite lattice with rank at most four can be primitively embedded in $E_8$, the unique $8$-dimensional positive-definite even unimodular lattice. Furthermore, every automorphism $g\in  SO_0(\Lambda)$ can be extended to an automorphism of $E_8$ that fixes the orthogonal complement $\Lambda^\perp$ of $\Lambda$ in $E_8$. Therefore, $ SO_0(\Lambda)$ is a subgroup of the group  $G_1\subset W^+(E_8)$ that fixes the sublattice $\Lambda^\perp \subset E_8$. 

Conversely, suppose that $G_1$ is the subgroup of $W^+(E_8)$ that stabilises pointwise a sublattice $\Lambda^\perp\equiv (E_8)^{G_1}$ of rank at least $4$. Then, $G_1$ acts faithfully as a group of automorphisms of the orthogonal complement $\Lambda$ of $\Lambda^\perp$. Furthermore, since $G_1$ fixes $(\Lambda^\perp)^*/\Lambda^\perp$, by the gluing construction it must also fix the discriminant group $\Lambda^*/\Lambda$ of its orthogonal complement. We conclude that $G_1\cong SO_0(\Lambda)$.

\subsection{Characterization of the groups of symmetries}\label{s:characterize}

In this section, we  classify all possible groups $G_1$ satisfying the equivalent properties (3) and (4) of theorem 2.

\medskip

Let $\Lambda$ be an even positive definite lattice of rank $r\le 4$ and let $SO_0(\Lambda)$ the group of positive determinant automorphisms that act trivially on the discriminant group $\Lambda^*/\Lambda$. By theorem 2, there is a torus model such that $G_1\cong SO_0(\Lambda)$. We recall that $\Lambda$ can always be primitively embedded in the $E_8$ lattice and that every $g\in SO_0(\Lambda)$ extends to an automorphism $\hat g\in W^+(E_8)$ that fixes the orthogonal complement of $\Lambda$ \cite{Nikulin}. Thus, $\Lambda$ can always be regarded as a primitive sublattice of $E_8$ and we can label each $g\in SO_0(\Lambda)$ by the $W^+(E_8)$ conjugacy class of $\hat g$.\footnote{The embedding of $\Lambda$ into the $E_8$ lattice is not necessarily unique, so we are implicitly making a choice here. Our subsequent analysis shows that the assignment of a $W^+(E_8)$ class to each $g\in SO_0(\Lambda)$ is independent of such a choice.} The group $W^+(E_8)$ contains 11 conjugacy classes whose elements fix a sublattice of rank at least $4$  \cite{Atlas}, see table \ref{t:classes}.

\begin{table}[h]
\begin{center}
\begin{tabular}{|c|cccc|cccc|c|c|c|}
\hline
Class & 
$\eta_1$ & $ \eta_1^{-1}$ & $\eta_2$ & $\eta_2^{-1}$ & $\pm\zeta_L$ & $\pm\zeta_L^{-1}$ & $\pm\zeta_R$ & $\pm\zeta_R^{-1}$ &$o(\pm g_0)$ & rk & det\\
\hline
1A & 
1 & 1&  1& 1& $\pm 1$ & $\pm 1$ &   $\pm 1$  & $\pm 1$ & 1,2 & 0 & -\\
2B 
& $1$ & $1$ & $-1$ & $-1$ & $\pm i$ & $\mp i$ & $\pm i$ & $\mp i$ & 4 & 2& 4\\
3A 
& $1$ & $1$ & $e^{\frac{2\pi i}{3}}$ & $e^{-\frac{2\pi i}{3}}$ & $\pm e^{\frac{2\pi i}{3}}$ & $\pm e^{-\frac{2\pi i}{3}}$ & $\pm e^{\frac{2\pi i}{3}}$ & $\pm e^{-\frac{2\pi i}{3}}$ & 3,6 & 2& 3\\
\hline
2A& 
$-1$ & $-1$ & $-1$ & $-1$ & $\pm 1$ & $\pm 1$ & $\mp 1$& $\mp 1$ & 2 & 4& 16\\
2E& 
$-1$ & $-1$ & $-1$ & $-1$ & $\pm 1$ & $\pm 1$ & $\mp 1$& $\mp 1$ & 2 & 4& 16\\
3E & 
$e^{\frac{2\pi i}{3}}$ & $e^{-\frac{2\pi i}{3}}$  & $e^{\frac{2\pi i}{3}}$& $e^{-\frac{2\pi i}{3}}$  & $\pm e^{\frac{2\pi i}{3}}$ &  $\pm e^{-\frac{2\pi i}{3}}$ & $\pm 1$ & $\pm 1$ & 3,6 & 4& 9\\
4A 
& $i$ & $-i$  & $i$ & $-i$ & $\pm i$ & $\mp i$ & $\pm 1$ & $\pm 1$ & 4,4 & 4& 4\\
4C 
& $i$ & $-i$ & $-1$ & $-1$ & $\pm e^{\frac{\pi i}{4}}$& $\pm e^{-\frac{\pi i}{4}}$ & $\mp e^{\frac{\pi i}{4}}$ &$\mp e^{-\frac{\pi i}{4}}$ & 8& 4& 8\\
5A & 
$e^{\frac{2\pi i}{5}}$ & $e^{-\frac{2\pi i}{5}}$ & $e^{\frac{4\pi i}{5}}$ & $e^{-\frac{4\pi i}{5}}$ & $\pm e^{\frac{2\pi i}{5}}$ & $\pm e^{-\frac{2\pi i}{5}}$ & $\pm e^{\frac{4\pi i}{5}}$ & $\pm e^{-\frac{4\pi i}{5}}$ & 5,10 & 4& 5\\
6A & 
$e^{\frac{2\pi i}{6}}$ & $e^{-\frac{2\pi i}{6}}$ & $-1$ & $-1$ & $\pm e^{\frac{2\pi i}{3}}$ & $\pm e^{-\frac{2\pi i}{3}}$ & $\pm e^{\frac{2\pi i}{6}}$ & $\pm e^{\frac{-2\pi i}{6}}$ & 6 & 4& 4\\
6D & 
$e^{\frac{2\pi i}{3}}$ & $e^{-\frac{2\pi i}{3}}$ & $-1$ & $-1$ & $\pm e^{\frac{2\pi i}{12}}$ & $\pm e^{-\frac{2\pi i}{12}}$ & $\pm e^{\frac{10\pi i}{12}}$ & $\pm e^{-\frac{10\pi i}{12}}$ & 12 & 4& 12\\
\hline
\end{tabular}
\end{center}
\caption{\small Conjugacy classes of elements $\hat g\in W^+(E_8)$ that fix a sublattice of rank at least $4$. 
For each class, we list the the eigenvalues $\eta_1,\eta_2,\eta^{-1}_1,\eta^{-1}_2$ of the corresponding $g\in G_1$ on the space $\Pi_\textrm{even}^-$, the rank of $(\Gamma^{4,4}_\textrm{even})_{\langle g\rangle}$ and the determinant of $(1-g)$ on $(\Gamma^{4,4}_\textrm{even})_{\langle g\rangle}$. The corresponding symmetry $\pm g_0\in G_0$ of the torus model is determined up to a sign. We list the eigenvalues $\pm\zeta_L,\pm\zeta_R,\pm\zeta_L^{-1},\pm\zeta_R^{-1}$ (each with multiplicity $2$) for the action of $\pm g_0$ over the fermions $\psi^a,\tilde \psi^a$ and the orders $o(\pm g_0)$ of $g_0$ and $-g_0$.
}\label{t:classes}
\end{table}
For a given $g\in SO_0(\Lambda)$,  the conjugacy class of $\hat g$ in $W^+(E_8)$ can be determined, in most cases, by the eigenvalues of $g$ on $\Lambda$. The only exceptions are the classes 2A and 2E, that have the same eigenvalues. To distinguish these cases, it is sufficient to notice that $g\in SO_0(\Lambda)$ corresponds to the class 2A if and only if it is the square of some $h\in SO_0(\Lambda)$ of class 4A. 

Let $\lambda_1,\ldots,\lambda_r$ be a basis of generators for $\Lambda$ and $Q_{ij}:=\lambda_i\cdot \lambda_j$ be the corresponding Gram matrix. Then $g\in SO_0(\Lambda)$ acts by
$$ g(\lambda_i)=\sum_{j=1}^r g_{ij}\lambda_j\ ,\qquad i=1,\ldots,r\ ,
$$ where $g_{ij}$ is an integral matrix satisfying
$$ g^t\, Q\, g =Q\ .
$$ Let $u_1,\ldots, u_r$ be the dual basis of generators of $\Lambda^*$,  given by \beq u_i:=\sum_j Q^{-1}_{ij}\lambda_j\in \Lambda\otimes \QQ\ ,\ee so that $u_i\cdot \lambda_j=\delta_{ij}$. 
The action of $g$ extends to $\Lambda\otimes \QQ$ by linearity, so that 
$$ u_i-g(u_i)=\sum_j (Q^{-1}_{ij}\lambda_j-Q^{-1}_{ij} g(\lambda_j))= \sum_{j=1}^r (Q^{-1}(1-g))_{ij}\lambda_j \ ,\qquad i=1,\ldots,r\ ,
$$ and the condition that $g$ acts trivially on $\Lambda^*/\Lambda$ is equivalent to
\beq\label{condo} (Q^{-1}(1-g))_{ij}\in\ZZ\ ,\qquad \text{for all } i,j=1,\ldots,r\ .
\ee
In particular, 
\beq\label{conda} \frac{\det (1-g)}{\det Q}\in \ZZ\ ,
\ee which puts a non-trivial constraint on the possible Gram matrices $Q$, provided that $\rk (1-g)=r$. The determinant $\det(1-g)$ and the rank $\rk (1-g)$ for each possible conjugacy class in $W^+(E_8)$ are given in table \ref{t:classes}. For each $W^+(E_8)$-conjugacy class of $g$, there are only few isomorphism classes of lattices $\Lambda$ of rank $r=\rk(1-g)$ and satisfying  \eqref{conda} (see \cite{Nipp}). In the following sections, we consider a case by case analysis of these lattices. First of all, in section \ref{s:geomgroups} we consider the case where $G_1$ has no symmetry $g$ with $\rk (1-g)=4$. In this case, we argue that the whole group $G_0$ is induced by geometric automorphisms of the target space in a suitable geometric interpretation of the CFT as a non-linear sigma model on $T^4$.

\bigskip

The lattices $\Lambda$ admitting symmetries $g\in SO_0(\Lambda)$ with $\rk (1-g)=4$ and the  $W^+(E_8)$ conjugacy classes of the associated lifts $\hat g$ are given in table \ref{t:lattici}. The corresponding groups $G_1=SO_0(\Lambda)$ and their double covering $G_0$ cannot be described as automorphisms of the target space in any geometric interpretation of the model. A detailed description of the entries of this table is provided in section \ref{s:nongeom}.

Table \ref{t:lattici} is derived as follows. When $1-g$ has rank $\rk (1-g)=4=r$, then the lattice $\Lambda$ is uniquely determined, up to isomorphism, by the $W^+(E_8)$ conjugacy class of $\hat g$. Indeed, in this case, $\Lambda$ is the orthogonal complement in $E_8$ of the sublattice fixed by $\hat g$ and for any two elements $\hat g,\hat g'\in W^+(E_8)$ in the same conjugacy class, the corresponding lattices $\Lambda$ and $\Lambda'$ are related by a $W^+(E_8)$ transformation and are therefore isomorphic.

\begin{table}[h]
\begin{center}
\begin{tabular}{cccccc}
Lattice $\Lambda$ & $\det Q$ & $W^+(E_8)$ classes & $G_1\cong SO_0(\Lambda)$ & $G_0$ & $|G_0|$ \\ \hline\\[-13pt]
$D_4$ & 4 & 2A,\ 4A,\ 6A& $\pm [T_{12}\times_{C_3} T_{12}]$ & $\T_{24}\times_{C_3} \T_{24}$ & 192 \\
$A_4$ & 5 & 5A & $+ [I_{60}\times_{I_{60}} \bar I_{60}]$ & $\I_{120}\times_{\I_{120}}\bar\I_{120}$ & 120\\
$A_1\, A_3$ & 8 & 4C &$+ [O_{24}\times_{O_{24}} \bar O_{24}]$ & $\calO_{48}\times_{\calO_{48}} \bar\calO_{48}$ & 48 \\
$A_2^2$ & 9 & 3E& $+ [D_{6}(3)\times_{C_2} D_{6}(3)]$ & $\D_{12}(3)\times_{\C_4} \D_{12}(3)$ & 36 \\
$A_1^2\, A_2$ & 12 & 6D& $+ [D_{12}(6)\!\times_{\!D_{12}(6)}\! D_{12}(6)]$ & $\D_{24}(6)\!\times_{\!\D_{24}(6)}\! \D_{24}(6)$ & 24 \\
$A_1^4 $ & 16 & 2E& $\pm [D_{4}(2)\times_{D_{4}(2)} D_{4}(2)]$ & $\D_{8}(4)\times_{D_{4}(2)} \D_{8}(4)$ & 16
\end{tabular}\end{center}\caption{\small Lattices $\Lambda$ of rank $4$ admitting automorphisms $g\in SO_0(\Lambda)$ with $\rk(1-g)=4$. For each lattice, we give the determinant of the Gram matrix $Q$, the $W^+(E_8)$ classes of the elements $g$ with $\rk(1-g)=4$, the group $G_1\cong SO_0(\Lambda)$, its $\ZZ_2$-central extension $G_0=\ZZ_2.G_1$ and its order $|G_0|$. See section \ref{s:nongeom} and appendix \ref{s:discrSpin} for a description of the lattices and groups.}\label{t:lattici}
\end{table}
By table \ref{t:classes} and \eqref{conda}, the only possible values for $\det Q$ are $4,5,8,9,12,16$. When $\det Q\in \{4,5,8,9\}$,  there is a unique isomorphism class of lattices $\Lambda$ for each value of the determinant \cite{Nipp} and this determines the corresponding entries in table \ref{t:lattici}. When $\det Q=12$ or $\det Q=16$ there are two isomorphism classes of lattices. The analysis of $SO_0(\Lambda)$ in section \ref{s:nongeom} shows that the lattices in the last two rows of table \ref{t:lattici} admit a symmetry in the corresponding $W^+(E_8)$ class. As argued above, each of the $W^+(E_8)$ classes in table \ref{t:classes} with $\rk(1-g)=4$ is associated with exactly one isomorphism class of lattices and table \ref{t:lattici} contains all such $W^+(E_8)$ classes. This implies that the lattices $\Lambda$ in the two isomorphism classes with $\det Q=12$ and $\det Q=16$ that do not appear in table  \ref{t:lattici} have no automorphism $g\in SO_0(\Lambda)$ with $\rk(1-g)=4$.

Finally, we argue that each of the six lattices in table \ref{t:lattici} corresponds to a unique point in the moduli space $\M_{T^4}$. In other words, there is a unique torus model (up to dualities) realizing each of the symmetry groups of table \ref{t:lattici}. In fact, the group $G_1$ of a torus model is isomorphic to $SO_0(\Lambda)$, with $\Lambda$ one of the entries of table \ref{t:lattici}, if and only if the lattice $S:=\Gamma^{4,4}_\textrm{even}\cap \Pi^-_\textrm{even}$ is isomorphic to $\Lambda(-1)$. Vice versa, each sublattice $S\subset \Gamma^{4,4}_\textrm{even}$ isomorphic to $\Lambda(-1)$ is associated with a unique torus model with symmetry group $G_1\cong SO_0(\Lambda)$, defined by setting $\Pi^-_\textrm{even}=S\otimes \RR$.  
Thus, we have to show that any two sublattices $S, S'\subset \Gamma^{4,4}_\textrm{even}$ with $S\cong \Lambda(-1)\cong S'$ are related by a $O(\Gamma^{4,4}_\textrm{even})$ transformation, so that the associated torus models are related by some T-duality. Let $K$ and $K'$ be the orthogonal complements in $\Gamma^{4,4}_\textrm{even}$ of $S$ and $S'$, respectively. Then, by the standard gluing construction, there are isomorphisms $\gamma:S^*/S\to K^*/K$  and $\gamma':(S')^*/S'\to (K')^*/K'$ between the discriminant groups such that the induced discriminant forms are inverted (see appendix \ref{s:lattices})
\beq\label{inve}  q_S=-q_K\circ \gamma\ ,\qquad q_{S'}=-q_{K'}\circ \gamma'\ ,
\ee and such that the lattice $\Gamma^{4,4}_\textrm{even}$ is given by
\beq\label{glue} \Gamma^{4,4}_\textrm{even}=\{(v,w)\in S^*\oplus K^*\mid \gamma([v])=[w]\}=\{(v,w)\in (S')^*\oplus (K')^*\mid \gamma'([v])=[w]\}\ ,
\ee where $[v]$ and $[w]$ are the images of $v$ and $w$ in the corresponding discriminant groups. The existence of the isomorphisms $\gamma$ and $\gamma'$ implies, in particular, that $K$ and $K'$ must be positive definite lattices of rank $4$ with the same determinant as $\Lambda$. In all six cases of table \ref{t:lattici}, one finds that $\gamma,\gamma'$ exist if and only if $K\cong \Lambda\cong K'$ \cite{Nipp}. Furthermore, the isomorphisms $\gamma$ and $\gamma'$ satisfying \eqref{inve} are unique up to automorphisms of $S$, $S'$, $K$, $K'$. It follows that we can always find isomorphisms $f_S:S^* \to (S')^*$ and $f_K: K^*\to  (K')^*$ such that the induced isomorphisms of discriminant groups satisfy
$$ f_K\circ \gamma = \gamma'\circ f_S\ .
$$ By \eqref{glue}, this implies that $f_S\oplus f_K:S^*\oplus K^*  \to (S')^*\oplus  (K')^*$ restricts to an automorphism of $\Gamma^{4,4}_\textrm{even}$ and, by construction, this automorphism maps $S$ to $S'$.

\subsection{Models with purely geometric symmetry group}\label{s:geomgroups}

In this section, we will show that $G_1$ contains no elements with $\rk(1-g)=4$ if and only if the whole group $G_0=\ZZ_2.G_1$ is induced by target space automorphisms in a suitable geometric interpretation of the torus model. We will also classify such purely geometric groups; analogous results were obtained in \cite{Wendland:2000ry}.

Suppose that $G_1\subset SO^+(\Gamma^{4,4}_\textnormal{even})$ contains only elements $g$ in class 1A, 2B or 3A of $W^+(E_8)$, so that $\rk(1-g)<4$ (see table \ref{t:classes}). Let $\Gamma^{G_1}\subset \Gamma^{4,4}_\textnormal{even}$ be the sublattice fixed by $G_1$ and $\Gamma_{G_1}$ its orthogonal complement. The lattice  $\Lambda:=\Gamma_{G_1}(-1)$  is an even positive definite lattice of rank at most $4$ and such that $SO_0(\Lambda)\cong G_1$; furthermore, there is no vector of $\Lambda$ fixed by all elements of $G_1$. We will show that under the above condition for $G_1$, $\Lambda$ has rank at most three. 

Let $\Lambda^{(2)}$ denote the sublattice of $\Lambda$ generated by all its elements of squared length $2$ (roots). Suppose first, by absurd, that $\Lambda^{(2)}$ has rank four. All root lattices of rank $4$ are listed in Table \ref{t:lattici}, so that $\Lambda^{(2)}$ must be one of them. Now, $\Lambda$ has the same rank as $\Lambda^{(2)}$ and its vectors have integral scalar product with all elements of $\Lambda^{(2)}$ so that $\Lambda^{(2)} \subseteq \Lambda\subseteq (\Lambda^{(2)})^*$. Furthermore, all the vectors of $\Lambda$ of squared norm two are contained in $\Lambda^{(2)}$. For all root lattices $\Lambda^{(2)}$ in table \ref{t:lattici}, the only sublattices of  $(\Lambda^{(2)})^*$ satisfying these properties are the root lattices themselves, so that necessarily $\Lambda=\Lambda^{(2)}$. But this contradicts the assumption that $G_1$ has only elements in the classes 1A, 2B and 3A. It follows that $\Lambda^{(2)}$ has rank at most three and we denote  by  $\hat \Lambda$ its orthogonal complement in $\Lambda$.

For each $g\in SO_0(\Lambda)$, $g\neq 1$,  let $\Lambda^{g}\subset \Lambda$ be the sublattice fixed by $g$ and $\Lambda_g$ its orthogonal complement in $\Lambda$.
By eq.\eqref{conda} and table \ref{t:classes}, $\Lambda_g$ has rank $\rk(1-g)=2$ and determinant $3$ (for $g$ in class 3A) or $4$ (for $g$ in class 2B), so that the only possibilities for the Gram matrix of $\Lambda_g$, up to isomorphism, are
$$ Q_{A_2}=\begin{pmatrix}
2 & -1\\ -1 & 2
\end{pmatrix}\ ,\qquad Q_{A_1^2}=\begin{pmatrix}
2 & 0\\ 0 & 2
\end{pmatrix}\ .
$$ 
In both cases, the lattice $\Lambda_g$ is generated by its elements of squared length $2$, so that $\Lambda_g\subseteq \Lambda^{(2)}$ for all $g\in G_1$. It follows that, for all $g\in G_1$, $\hat \Lambda$ is orthogonal to $\Lambda_g$, so that it must be fixed by $g$. However, by construction, there is no sublattice fixed by all $g\in G_1$, so that $\hat \Lambda=0$ and $\Lambda$ has the same rank as $\Lambda^{(2)}$ and thus at most three.

Since $\Gamma_{G_1}=\Lambda(-1)$ is a primitive sublattice of $\Gamma^{4,4}_\textnormal{even}$ with rank  $r\le 3$, then by Corollary 1.13.4 of \cite{Nikulin} the orthogonal complement $\Gamma^{G_1}$ is isomorphic to the orthogonal sum $\Gamma^{s,s}\oplus \Lambda$,  where $\Gamma^{s,s}$ is the even unimodular lattice of signature $(s,s)$, with $s:=4-r\ge 1$.
Therefore, if $G_1$ contains no elements $g$ with $\rk(1-g)=4$,   then there exists a pair of primitive null vectors  $\lambda_1,\lambda_2\in \Gamma^{4,4}_\textnormal{even}$ with $\lambda_1 \bullet \lambda_2=1$ that are fixed by $G_1$. As described in section \ref{s:Dbtrial},  two such elements of $\Gamma^{4,4}_\textnormal{even}$ correspond to two maximal isotropic sublattices $\Lambda_1,\Lambda_2\subset \Gamma^{4,4}_\textnormal{w--m}$, with $\Lambda_1\oplus\Lambda_2\cong \Gamma^{4,4}_\textnormal{w--m}$, that are fixed (setwise) by $G_0=\ZZ_2.G_1$. In turn, $\Lambda_1,\Lambda_2$ define a geometric interpretation of the CFT as a non-linear sigma model on some torus $T^4=\RR^4/L$ (see section \ref{s:generalities}), such that  the lattice $L$ and the B-field are preserved by $G_0$.  With respect to this geometric interpretation, $G_0$ acts by $SU(2)$ transformations \eqref{second},\eqref{last} with $g_L=g_R$ on the complex coordinates $Z^{(1)},Z^{(2)}$ on $T^4$ defined by \eqref{zdef}.

\medskip

The groups of automorphisms of complex tori of complex dimension $2$ have been studied by Fujiki \cite{Fujiki}. Upon a suitable choice of the B-field, each such automorphism induces a symmetry of the corresponding non-linear sigma model, preserving a $\N=(4,4)$ superconformal algebra \cite{Wendland:2000ry}. In particular, $G_0$ corresponds to the group of fixed-point automorphisms of the torus. In \cite{Fujiki}, the action of these geometric automorphisms on the integral second cohomology group $H^2(T^4,\ZZ)$ is considered; this is the exact geometric  analogue of the action of $G_1=G_0/(-1)$ on even D-brane charges. More precisely, for a given geometric interpretation of the CFT, the lattice $\Gamma^{4,4}_\textnormal{even}$ can be identified with the even integral cohomology group $H^\textnormal{even}(T^4,\ZZ)$. In the interpretation where $G_1$ is purely geometric, the components of $H^\textnormal{even}(T^4,\ZZ)$ of degrees zero and four are fixed, so that $(\Gamma^{4,4}_\textnormal{even})_{G_1}$ can be identified with $H^2(T^4,\ZZ)_{G_1}$. Our analysis, therefore, naturally leads to the same result as in Lemma 3.3 and Theorem 6.4 of \cite{Fujiki}. The list of possible lattices $\Lambda=(\Gamma^{4,4}_\textnormal{even})_{G_1}(-1)$, their rank, the group $G_1\cong SO_0(\Lambda)$ and its $\ZZ_2$-central extension $G_0$ are given in table \ref{t:geom}.
\begin{table}[h!]
\begin{center}\begin{tabular}{cccccc} 
$\Lambda$ & Rank & $G_1$ & $G_0$ & $|G_0|$ & Generators of $G_0$\\
\hline
$0$ & 0 & $1$ & $\C_2$ & 2 & $(-1,-1)$\\
$A_1^2$ & 2 & $\ZZ_2$ & $\C_4$ & 4&  $(\ii,\ii)$\\
$A_2$ & 2 & $\ZZ_3$ & $\C_6$ & 6 & $(\eei_{\frac{1}{6}},\eei_{\frac{1}{6}})$ \\
$A_1^3$ & 3 & $\ZZ_2^2$ & $\D_8(2)$ & 8 & $(\ii,\ii),\quad (\jj,\jj)$ \\
$A_1\, A_2$ & 3 & $\ZZ_3\times \ZZ_2$ & $\D_{12}(3)$ & 12 &$(\jj,\jj),\quad (\eei_{\frac{1}{6}},\eei_{\frac{1}{6}})$ \\
$A_3$ & 3 & $Alt(4)$ & $\T_{24}$ & 24 & $(\ii,\ii), \quad (\om,\om)$\\[-15pt]
\end{tabular}\end{center}\caption{\small
The possible lattices $\Lambda\cong (\Gamma^{4,4}_\textrm{even})_{G_1}$ with $\rk\Lambda<4$. For each such lattice, we report the rank, the group $G_1=SO_0(\Lambda)$ of automorphisms fixing the discriminant group $\Lambda^*/\Lambda$, the double covering $G_0=\ZZ_2.G_1$ and the order and generators of $G_0\subset SU(2)\times SU(2)$. We use the notation $\eei_{r}:=\exp(2\pi\ii r)$, $r\in\QQ$, and $\om=\frac{1}{2}(-1+\ii+\jj+\kk)$.}\label{t:geom}\end{table}
Notice that, since $\Lambda$ has rank at most three, $G_1$ is actually a subgroup of $SO(3)$ and its spin cover $G_0$ is a subgroup of $SU(2)$. This $SU(2)$ is diagonally embedded in the group $SU(2)_L\times SU(2)_R$ acting as in \eqref{first}--\eqref{last}, that is $G_0\subset SU(2)_L\times SU(2)_R$ is generated by elements of the form $(g_L,g_R)=(g,g)$ for some $g\in SU(2)$.

The torus models with a given symmetry group $G_1$, associated with a lattice $\Lambda=(\Gamma^{4,4}_\textnormal{even})_{G_1}(-1)$ of rank $r$, form a sublocus of dimension $4(4-r)$ in the moduli space. Indeed, this is the dimension of the Grassmannian parametrizing the choice of a four dimensional positive-definite space $\Pi^+_\textnormal{even}$ within the space $(\Gamma^{4,4}_\textnormal{even})^{G_1}\otimes \RR$ of signature $(4,4-r)$. 

\subsection{Models with non-geometric symmetries}\label{s:nongeom}

As discussed in section \ref{s:characterize}, there are six torus models that admit symmetries preserving the $\N=(4,4)$ algebra and with no geometric interpretation. They are characterized by the condition that $\Gamma^{4,4}_\textrm{even}\cap \Pi^-_\textrm{even}\cong \Lambda(-1)$, where $\Lambda$ is one of the six lattices in table \ref{t:lattici}. In this section, we provide some more details about the lattices $\Lambda$, the groups $G_1=SO_0(\Lambda)$ and the corresponding torus models.

For each of these six cases, we first describe the lattice $\Lambda$: we provide the matrix $Q_{ij}:=\lambda_i\cdot\lambda_j$ for a standard choice of generators $\lambda_1,\ldots,\lambda_4$ of $\Lambda$, the determinant $\det Q$, the structure of the discriminant group $\Lambda^*/\Lambda$ and the values of the induced discriminant form $q:\Lambda^*/\Lambda\to \QQ/2\ZZ$ on its generators. We also provide a definition of $\Lambda\subset \RR^4$ as a lattice in the space of quaternions $\HH\cong \RR^4$. Next, we describe the group $G_1\cong SO_0(\Lambda)$ of automorphisms of $\Lambda$ with positive determinant and that act trivially on the discriminant group. In all the cases we are interested in, $\Lambda$ is the root lattice of some Lie algebra and $SO_0(\Lambda)$ is the group $W^+(\Lambda)$ of even Weyl transformations.  Since $G_1$ is a finite subgroup of $SO(4)\cong (SU(2)\times SU(2))/(-1,-1)$, it can be described as $G_1=\pm [L\times_F R]$ or $G_1=+ [L\times_F R]$, depending on whether $-1\in G_1$ or not (see appendix \ref{s:discrSpin}). We classify the elements of $G_1$ depending on their order and conjugacy class in $W^+(E_8)$ and for each such class we indicate the number of elements in $G_1$.  

The group $G_0$ is the preimage of $G_1$ under the covering map $\Spin(4)\to SO(4)$ and it can be easily determined as described in appendix \ref{s:discrSpin}. We provide a set of generators $(g_L,g_R)$ of $G_0\subset SU(2)\times SU(2)$ in the form of a pair of quaternions. The corresponding generators of $G_1$ are the $SO(4)$ transformations $[g_L,g_R]\equiv [-g_L,-g_R]$ acting by
$$ [g_L,g_R]:\lambda\mapsto g_L\, \lambda\, \bar g_R\ ,\qquad\qquad \lambda\in \Lambda\subset \HH\ ,\quad [g_L,g_R]\in G_1\ ,
$$ on the lattice $\Lambda=(\Gamma^{4,4}_\textnormal{even})_{G_1}(-1)$.

Next, we provide a description of the winding-momentum lattice $\Gamma^{4,4}_\textnormal{w--m}$ as a lattice in $\RR^{4,4}\cong \Pi_L\oplus \Pi_R\cong \HH\oplus \HH $. In other words, we denote the elements of $\Gamma^{4,4}_\textnormal{w--m}$ as pairs of quaternions $\ (q_L;q_R)\in \Gamma^{4,4}_\textnormal{w--m}$, where $q_L\in \Pi_L\cong \HH$, $q_R\in \Pi_R\cong \HH$, so that the action of $(g_L,g_R)\in G_0\subset SU(2)\times SU(2)$ is simply given by left multiplication
$$ (g_L,g_R):(q_L;q_R)\mapsto (g_Lq_L;g_Rq_R)\qquad\qquad (q_L,q_R)\in \Gamma^{4,4}_\textnormal{w--m},\quad (g_L,g_R)\in G_0\ ,
$$ as in eq.\eqref{mezzo}. The action of $(g_L,g_R)$ on the fields of the theory is given by \eqref{first}--\eqref{last}. We also describe the sublattices $\Gamma^{4,4}_\textrm{w--m}\cap \Pi_L$ and $\Gamma^{4,4}_\textrm{w--m}\cap \Pi_R$ of purely holomorphic and purely anti-holomorphic winding-momenta. If these sublattices have positive rank, then the chiral algebra of the model is extended with respect to the one of a generic torus model. 

Finally, we provide a possible geometric interpretation of the CFT as a non-linear sigma model with target space $\RR^4/L$ and B-field $B$. The information about $L$ and $B$ is given in the form of a matrix whose columns $l_1,\ldots,l_4$ form a set generators for $L\subset \RR^4$, the quadratic from $Q_{ij}=l_i\cdot l_j$ and the B-field $B$ in the form of a real antisymmetric matrix, such that
$$ \Gamma^{4,4}_\textrm{w--m}=\bigl\{\frac{1}{\sqrt{2}} \bigl(n_il^{i*}+w^il_i+w^i B_{ij}l^{j*}\,;\,n_il^{i*}-w^il_i+w^iB_{ij}l^{j*}\bigr)\mid n_1,\ldots,n_4,w^1,\ldots w^4\in \ZZ\bigr\}\ .
$$ Here $l^{1*},\ldots,l^{4*}$ is the basis of $L^*$ dual to $l_1,\ldots, l_4$, given by $l^{i*}=(Q^{-1})^{ij} l_j$.

\newpage

\subsubsection{Lattice $D_4$%
}\label{s:D4}

$\Lambda_{D_4}$ is the root lattice of $so(8)$
$$ Q_{D_4}=\begin{pmatrix}
 2  & -1 & 0 & 0\\ 
 -1 & 2 & -1 & -1 \\
 0 & -1 &2 & 0 \\
 0 &-1 & 0 &2
\end{pmatrix} \qquad\qquad  \begin{matrix}
\det Q=4\\[8pt]
\Lambda^*/\Lambda=\langle x,y\rangle\cong \ZZ_2\times\ZZ_2\\[8pt]
q(x)=q(y)=q(x+y)=1\mod 2\ZZ
\end{matrix}
$$
Quaternionic:
$$ \Lambda_{D_4}\cong \{a+b\ii+c\jj+d\kk\in \HH \mid a,b,c,d\in \ZZ, a+b+c+d\in 2\ZZ\}
$$
$$ \Lambda_{D_4}^*\cong \{a+b\ii+c\jj+d\kk\in \HH \mid (a,b,c,d)\in \ZZ^4\text{ or } (\ZZ+\tfrac{1}{2})^4\}\cong \frac{1}{\sqrt{2}}\Lambda_{D_4}\ ,
$$ 

\bigskip

\noindent Automorphism group $G_1=SO_0(\Lambda_{D_4})$:
$$ \begin{matrix}
G_1=\pm[T_{12}\times_{C_3} T_{12}]\\[8pt]
|G_1|=96
\end{matrix} \qquad \qquad \qquad 
\begin{array}{r|cccccc}
\text{Order } & 1 & 2 & 2&  3 & 4 & 6\\ W^+(E_8)\text{ class } & 1A & 2A & 2B &  3A & 4A & 6A
\\ \# \text{ elements } & 1 & 1 & 18 & 32 & 12 &32
\end{array}
$$

\bigskip

\noindent Symmetry group $G_0= \T_{24}\times_{C_3}\T_{24}$, order $|G_0|=192$,  with generators
$$  G_0=\langle (1,\ii),\ (1,\jj),\ (\ii,1),\ (\jj,1),\ (\om,\om)\rangle\qquad \qquad \om=\frac{1}{2}(-1+\ii+\jj+\kk)
$$

\bigskip

\noindent Lattice of winding-momenta:
\begin{multline*}
\Gamma^{4,4}_\textnormal{w--m}=\left\{\frac{1}{\sqrt{2}}(a_1+a_2\,\ii+a_3\,\jj+a_4\,\kk;\ b_1+b_2\,\ii+b_3\,\jj+b_4\,\kk)\mid a_i,b_i\in\ZZ,\right.\\ \left.\sum_ia_i\in 2\ZZ,\  a_1-b_1\equiv a_2-b_2\equiv a_3-b_3\equiv a_4-b_4 \mod 2\right\}
\end{multline*}

\medskip

\noindent Purely left- and purely right-moving momenta:
$$\Gamma^{4,4}_\textnormal{w--m}\cap \Pi_L\cong \Gamma^{4,4}_\textnormal{w--m}\cap \Pi_R=\left\{\sqrt{2}(a_1+a_2\,\ii+a_3\,\jj+a_4\,\kk)\mid (a_1,\ldots,a_4)\in \ZZ^4\text{ or }(\ZZ+\tfrac{1}{2})^4\right \}\cong \Lambda_{D_4}$$

\medskip

\noindent Geometric description as torus model on $\RR^4/L$ and B-field $B$
$$L=\begin{pmatrix}
1 & 0 & 0 & 0 \\
 -1 & 1 & 0 & 0 \\
 0 & -1 & 1 & 1 \\
 0 & 0 & -1 & 1
\end{pmatrix}
\qquad 
 Q_L=\begin{pmatrix}
 2  & -1 & 0 & 0\\ 
 -1 & 2 & -1 & -1 \\
 0 & -1 &2 & 0 \\
 0 &-1 & 0 &2
\end{pmatrix}\qquad\qquad  B= \begin{pmatrix}
 0 & 0 & -\frac{1}{2} & -\frac{1}{2} \\
 0 & 0 & 0 & 0 \\
 \frac{1}{2} & 0 & 0 & -\frac{1}{2} \\
 \frac{1}{2} & 0 & \frac{1}{2} & 0
\end{pmatrix}
$$

\newpage

\subsubsection{Lattice $A_4$
}\label{s:A4}

$\Lambda_{A_4}$ is the root lattice of $su(5)$
$$ Q_{A_4}=\begin{pmatrix}
 2  & -1 & 0 & 0\\ 
 -1 & 2 & -1 & 0 \\
 0 & -1 &2 & -1 \\
 0 & 0 & -1 &2
\end{pmatrix} \qquad\qquad  \begin{matrix}
\det Q=5\\[8pt]
\Lambda^*/\Lambda=\langle x\rangle\cong \ZZ_5\\[8pt]
q(x)=q(4x)= \frac{4}{5},\quad 
q(2x)=q(3x)= \frac{6}{5}
\end{matrix}
$$
Quaternionic:
$$\Lambda_{A_4}=\{\sqrt{2}(a+b\,\ii+c\,\om +d\, \iiI)\mid a,b,c,d\in\ZZ \}\qquad  \om=\tfrac{1}{2}(-1+\ii+\jj+\kk),\quad  \iiI=\tfrac{1}{2}(\ii+\tfrac{-\sqrt{5}-1}{2}\jj+\tfrac{\sqrt{5}-1}{2}\kk)$$

\bigskip

\noindent Automorphism group $G_1=SO_0(\Lambda_{A_4})$:
$$ \begin{matrix}
G_1= +[I_{60}\times_{I_{60}} \bar I_{60}]\cong Alt(5)\\[8pt]
|G_1|=60
\end{matrix} \qquad \qquad \qquad 
\begin{array}{r|cccc}
\text{Order } & 1 & 2 & 3 & 5\\ W^+(E_8)\text{ class } & 1A & 2B & 3A & 5A
\\ \# \text{ elements } &  1 & 15 & 20 &24
\end{array}
$$

\bigskip

\noindent Symmetry group $G_0= \I_{120}\times_{\I_{120}}\bar\I_{120}\cong 2.Alt(5)$, order $|G_0|=120$,  with generators
$$  G_0=\bigl\langle \ (\om,\eta(\om)),\quad (\iiI,\eta(\iiI))\ \bigr\rangle,\qquad\qquad
 \om=\tfrac{1}{2}(-1+\ii+\jj+\kk),\quad  \iiI=\tfrac{1}{2}(\ii+\tfrac{-\sqrt{5}-1}{2}\jj+\tfrac{\sqrt{5}-1}{2}\kk)
$$
The $\QQ$-linear map $\eta$ acts by $\sqrt{5}\mapsto -\sqrt{5}$, $\ii\mapsto -\ii$, $\jj\mapsto -\kk$, $\kk\mapsto -\jj$. It is an outer automorphism of $\I_{120}$. In particular, $\eta(\om)=\bar\om$ and $\eta(\iiI)=\bar{\boldsymbol{\iota}}_I$.

\bigskip

\noindent Lattice of winding-momenta:
\begin{multline*}
\Gamma^{4,4}_\textnormal{w--m}=\frac{\sqrt{2}}{5^{1/4}}\;\bigl\langle (1;1),\ (\ii;\eta(\ii)),\ (\kk;\eta(\kk)),\ (\om;\eta(\om)),\\ (\iiI;\eta(\iiI)),\ (\ii\iiI;\eta(\ii\iiI)),\ (\jj\iiI;\eta(\jj\iiI)),\ (\om\iiI;\eta(\om\iiI))\bigr\rangle_{\ZZ}
\end{multline*}

\bigskip

\noindent Purely left- and purely right-moving momenta:
$\Gamma^{4,4}_\textnormal{w--m}\cap \Pi_L =0= \Gamma^{4,4}_\textnormal{w--m}\cap \Pi_R$

\bigskip

\noindent Geometric description as torus model on $\RR^4/L$ and B-field $B$
$$L=\frac{\sqrt{2}}{5^{1/4}}
\begin{pmatrix}
1 & 0 & 0 & 0 \\
 -1 & 1 & 0 & 0 \\
 0 & -1 & 1 & 1 \\
 0 & 0 & -1 & 1
\end{pmatrix}
\qquad 
 Q_L=\frac{2}{\sqrt{5}}\begin{pmatrix}
 2  & -1 & 0 & 0\\ 
 -1 & 2 & -1 & -1 \\
 0 & -1 &2 & 0 \\
 0 &-1 & 0 &2
\end{pmatrix}\qquad\qquad  B= \begin{pmatrix}
0 & -\frac{1}{2} & -\frac{3}{4} & -\frac{1}{4} \\
 \frac{1}{2} & 0 & -\frac{1}{2} & -\frac{1}{2} \\
 \frac{3}{4} & \frac{1}{2} & 0 & \frac{1}{4} \\
 \frac{1}{4} & \frac{1}{2} & -\frac{1}{4} & 0
\end{pmatrix}
$$

\newpage

\subsubsection{Lattice $A_1A_3$ 
}\label{s:A1A3}

$\Lambda_{A_1\, A_3}$ is the root lattice of $su(2)\oplus su(4)$
$$ Q_{A_1\, A_3}=\begin{pmatrix}
 2  & 0 & 0 & 0\\ 
 0 & 2 & -1 & 0 \\
 0 & -1 &2 & -1 \\
 0 & 0 & -1 &2
\end{pmatrix} \qquad\qquad  \begin{matrix}
\det Q=8\\[8pt]
\Lambda^*/\Lambda=\langle x,y\rangle\cong \ZZ_2\times \ZZ_4\\[8pt]
q(y)=q(3y)=\frac{3}{4},\quad q(2y)=1,\quad
q(ax+by)=\frac{a^2}{2}+q(by)
\end{matrix}
$$
Quaternionic:
$$\Lambda_{A_1\, A_3}=\{(a+b\ii+c\jj+d\kk)\om+\sqrt{2}e\mid a+b+c+d=0, \ \ a,b,c,d,e\in \ZZ\}$$
\bigskip

\noindent Automorphism group $G_1=SO_0(\Lambda_{A_1\, A_3})$:
$$ \begin{matrix}
G_1\cong +[O_{24}\times_{O_{24}}\bar O_{24}] \cong Sym(4)\\[8pt]
|G_1|=24
\end{matrix} \qquad \qquad \qquad 
\begin{array}{r|cccc}
\text{Order } & 1 & 2 & 3 & 4\\ W^+(E_8)\text{ class } & 1A & 2B & 3A & 4C
\\ \# \text{ elements } &  1 & 9 & 8 & 6
\end{array}
$$

\bigskip

\noindent Symmetry group $G_0= \calO_{48}\times_{\!\calO_{48}}\bar{\calO}_{48}\cong 2.Sym(4)$, order $|G_0|=48$,  with generators
$$   G_0=\langle\, (\om,\om),\quad (\iiO,-\iiO)\,\rangle  ,
\qquad\qquad  \om=\tfrac{1}{2}(-1+\ii+\jj+\kk),\quad \iiO=\frac{\jj+\kk}{\sqrt{2}}
$$

\medskip

\noindent Lattice of winding momenta:
\begin{multline*}
\Gamma^{4,4}_\textnormal{w--m}=2^{-\frac{1}{4}}\;\bigl\langle (1;1),\ (\ii;\ii),\ (\jj;\jj),\ (\om;\om),\ (\iiO;-\iiO),\ (\ii \iiO;-\ii \iiO),\ (\jj \iiO;-\jj \iiO),\ (\kk \iiO;-\kk \iiO)\bigr\rangle_{\ZZ}
\end{multline*}

\medskip

\noindent Purely left- and purely right-moving momenta:
$\Gamma^{4,4}_\textnormal{w--m}\cap \Pi_L =0= \Gamma^{4,4}_\textnormal{w--m}\cap \Pi_R$

\bigskip

\noindent Geometric description as torus model on $\RR^4/L$ and B-field $B$
$$L=\frac{1}{2^{3/4}}
\begin{pmatrix}
2 & -1 & 0 & 0 \\
 0 & \sqrt{3} & 0 & 0 \\
 0 & 0 & 2 & -1 \\
 0 & 0 & 0 & \sqrt{3} 
\end{pmatrix}
\qquad 
 Q_L=\frac{1}{\sqrt{2}}\begin{pmatrix}
 2  & -1 & 0 & 0\\ 
 -1 & 2 &  0 & 0 \\
 0 & 0 &2 & -1 \\
 0 &0  & -1&2
\end{pmatrix}\qquad\qquad  B= \begin{pmatrix}
0 & 0 & \frac{1}{3} & -\frac{1}{3} \\
 0 & 0 & -\frac{1}{3} & \frac{1}{3} \\
 -\frac{1}{3} & \frac{1}{3} & 0 & 0 \\
 \frac{1}{3} & -\frac{1}{3} & 0 & 0
\end{pmatrix}
$$

\newpage

\subsubsection{Lattice $A_2^2$
}\label{s:A2A2}

$\Lambda_{A_2^2}$ is the root lattice of $su(3)\oplus su(3)$
$$ Q_{A_2^2}=\begin{pmatrix}
 2  & -1 & 0 & 0\\ 
 -1 & 2 & 0 & 0 \\
 0 & 0 &2 & -1 \\
 0 & 0 & -1 &2
\end{pmatrix} \qquad\qquad  \begin{matrix}
\det Q=9\\[8pt]
\Lambda^*/\Lambda=\langle x,y\rangle\cong \ZZ_3\times \ZZ_3\\[8pt]
q(x)=q(2x)=q(y)=q(2y)=\frac{2}{3},\\[3pt]
q(ax+by)=q(ax)+q(by)
\end{matrix}
$$
Quaternionic:
$$\Lambda_{A_2^2}=\{\sqrt{2}(a+b\,\eei_{\frac{1}{3}}+c\,\jj+d\,\jj\, \eei_{\frac{1}{3}})\mid a,b,c,d\in \ZZ\}\qquad \eei_{\frac{1}{3}}:=\exp(2\pi \ii/3)$$
\bigskip

\noindent Automorphism group $G_1=SO_0(\Lambda_{A_2^2})$:
$$ \begin{matrix}
G_1\cong +[D_6(3)\times_{C_2} D_6(3)]\cong (\ZZ_3\times \ZZ_3).\ZZ_2\\[8pt]
|G_1|=18
\end{matrix} \qquad \qquad \qquad 
\begin{array}{r|cccc}
\text{Order } & 1 & 2 & 3 & 3 \\ W^+(E_8)\text{ class } & 1A & 2B & 3A & 3E
\\ \# \text{ elements } &  1 & 9 & 4 & 4
\end{array}
$$

\bigskip

\noindent Symmetry group $\displaystyle{G_0= \D_{12}(3)\times_{\C_4}\D_{12}(3)}$, order $|G_0|=36$,  with generators
$$  G_0=\langle\ (1,\eei_{\frac{1}{3}}),\ (\eei_{\frac{1}{3}},1),\ (-\jj,\jj)\,\rangle\qquad\qquad \eei_{\frac{1}{3}}:=\exp(2\pi \ii/3)=-\frac{1}{2}+\frac{\sqrt{3}}{2}\ii
$$

\bigskip

\noindent Lattice of winding momenta:
\begin{multline*}
\Gamma^{4,4}_\textnormal{w--m}=\left\{\sqrt{\frac{2}{3}}(a_1+a_2\,\eei_{\frac{1}{3}}+b_1\,\jj+b_2\,\jj\, \eei_{\frac{1}{3}};\ \tilde a_1+\tilde a_2\,\eei_{\frac{1}{3}}+\tilde b_1\,\jj+\tilde b_2\,\jj\, \eei_{\frac{1}{3}})\mid a_i,b_i,\tilde a_i,\tilde b_i\in\ZZ,\right.\\\left. a_1+a_2\equiv \tilde a_1+\tilde a_2\mod 3, \quad b_1+b_2\equiv -(\tilde b_2+\tilde b_1)\mod 3 \right\}
\end{multline*}

\medskip

\noindent Purely left- and purely right-moving momenta:
$$\Gamma^{4,4}_\textnormal{w--m}\cap \Pi_{L,R}=\left\{ \sqrt{\tfrac{2}{3}}\bigl(a_1(1-\eei_{\frac{1}{3}})+3a_2+b_1\,\jj(1-\eei_{\frac{1}{3}})+3b_2\,\jj\bigr) \mid a_1,a_2,b_1,b_2\in\ZZ\right\}\cong \Lambda_{A_2^2}$$

\medskip 

\noindent Geometric description as torus model on $\RR^4/L$ and B-field $B$
$$L=\frac{2}{\sqrt{3}}
\begin{pmatrix}
1 & -\frac{1}{2} & 0 & 0 \\
 0 & \frac{\sqrt{3}}{2} & 0 & 0 \\
 0 & 0 & 1 & -\frac{1}{2} \\
 0 & 0 & 0 & \frac{\sqrt{3}}{2} 
\end{pmatrix}
\qquad 
 Q_L=\frac{2}{3}\begin{pmatrix}
 2  & -1 & 0 & 0\\ 
 -1 & 2 &  0 & 0 \\
 0 & 0 &2 & -1 \\
 0 &0  & -1&2
\end{pmatrix}\qquad\qquad  B= \begin{pmatrix}
0 & -\frac{2}{3} & 0 & 0 \\
 \frac{2}{3} & 0 & 0 & 0 \\
 0 & 0 & 0 & -\frac{2}{3} \\
 0 & 0 & \frac{2}{3} & 0
\end{pmatrix}
$$

\newpage

\subsubsection{Lattice $A_1^2 A_2$
}\label{s:A1A1A2}

$\Lambda_{A_1^2\, A_2}$ is the root lattice of $su(2)\oplus su(2)\oplus su(3)$
$$ Q_{A_1^2\, A_2}=\begin{pmatrix}
 2  & 0 & 0 & 0\\ 
 0 & 2 & 0 & 0 \\
 0 & 0 &2 & -1 \\
 0 & 0 & -1 &2
\end{pmatrix} \qquad\qquad  \begin{matrix}
\det Q=12\\[8pt]
\Lambda^*/\Lambda=\langle x,y,z\rangle\cong \ZZ_2\times\ZZ_2\times \ZZ_3\\[8pt]
q(x)=q(y)=1/2,\quad q(z)=q(2z)=\frac{2}{3},\\
q(ax+by+cz)=q(ax)+q(by)+q(cz)
\end{matrix}
$$
Quaternionic:
$$\Lambda_{A_1^2\, A_2}=\{\sqrt{2}(a+b\,\ii+c\,\jj+d\,\jj \,\eei_{\frac{1}{3}})\mid a,b,c,d\in \ZZ\}\qquad \eei_{\frac{1}{3}}=\exp(2\pi\ii/3)$$
\bigskip

\noindent Automorphism group $G_1=SO_0(\Lambda_{A_1^2\, A_2})$:
$$ \begin{matrix}
G_1= +[D_{12}(6)\times_{D_{12}(6)} D_{12}(6)]\\[3pt] \quad \cong D_{12}(6) \cong \ZZ_2\times Sym(3)\\[8pt]
|G_1|=12
\end{matrix} \qquad \qquad \qquad 
\begin{array}{r|cccc}
\text{Order } & 1 & 2 & 3 & 6 \\ W^+(E_8)\text{ class } & 1A & 2B & 3A & 6D 
\\ \# \text{ elements } &  1 & 7 & 2 & 2
\end{array}
$$

\bigskip

\noindent Symmetry group $G_0= \D_{24}(6)\times_{\D_{24}(6)} \D_{24}(6)\cong \D_{24}(6)$, order $|G_0|=24$,  with generators
$$ G_0=\langle\ (\eei_{\frac{1}{3}},\eei_{\frac{1}{3}}),\quad (-\jj,\jj),\quad (-\ii,\ii) \rangle
$$

\bigskip

\noindent Lattice of winding momenta:
\begin{multline*}
\Gamma^{4,4}_\textnormal{w--m}=
3^{-\frac{1}{4}}\Bigl(\bigl\langle (1;1),\ (\ii;-\ii),\ (\eei_{\frac{1}{3}};\eei_{\frac{1}{3}}),\ (\ii\, \eei_{\frac{1}{3}};-\ii\, \eei_{\frac{1}{3}})\bigr\rangle_{\ZZ}\\
\oplus \bigl\langle (\jj;-\jj),\ (\kk;\kk),\ (\jj\, \eei_{\frac{1}{3}};-\jj\, \eei_{\frac{1}{3}}),\ (\kk \,\eei_{\frac{1}{3}};\kk\, \eei_{\frac{1}{3}})\bigr\rangle_{\ZZ}\Bigr)
\end{multline*}

\bigskip

\noindent Purely left- and purely right-moving momenta:
$\Gamma^{4,4}_\textnormal{w--m}\cap \Pi_L =0= \Gamma^{4,4}_\textnormal{w--m}\cap \Pi_R$

\bigskip

\noindent Geometric description as torus model on $\RR^4/L$ and B-field $B$
$$L=\frac{\sqrt{2}}{3^{1/4}}
\begin{pmatrix}
1  & 0 & 0 & 0\\ 
 0 & 1 &  0 & 0 \\
 0 & 0 &1 & 0 \\
 0 &0  & 0&1 
\end{pmatrix}
\qquad 
 Q_L=\frac{2}{\sqrt{3}}\begin{pmatrix}
 1  & 0 & 0 & 0\\ 
 0 & 1 &  0 & 0 \\
 0 & 0 &1 & 0 \\
 0 &0  & 0&1
\end{pmatrix}\qquad\qquad  B= \begin{pmatrix}
0 & -\frac{1}{2} & 0 & 0 \\
 \frac{1}{2} & 0 & 0 & 0 \\
 0 & 0 & 0 & \frac{1}{2} \\
 0 & 0 & -\frac{1}{2} & 0
\end{pmatrix}
$$

\newpage

\subsubsection{Lattice $A_1^4$
}\label{s:4A1}

$\Lambda_{A_1^4}$ is the root lattice of $su(2)\oplus su(2)\oplus su(2)\oplus su(2)$
$$ Q_{A_1^4}=\begin{pmatrix}
 2  & 0 & 0 & 0\\ 
 0 & 2 & 0 & 0 \\
 0 & 0 &2 & 0 \\
 0 & 0 & 0 &2
\end{pmatrix} \qquad\qquad  \begin{matrix}
\det Q=16\\[8pt]
\Lambda^*/\Lambda=\langle x_1,\ldots,x_4\rangle\cong \ZZ_2^4\\[8pt]
q(\sum_i a_i x_i)=\frac{1}{2}\sum_i a_i, \qquad a_i\in \{0,1\}
\end{matrix}
$$
Quaternionic:
$$ \Lambda_{A_1^4}=\{\sqrt{2}(a+b\,\ii+c\,\jj+d\,\kk)\mid a,b,c,d\in\ZZ\}
$$
\bigskip

\noindent Automorphism group $G_1=SO_0(\Lambda_{A_1^4})$:
$$ \begin{matrix}
G_1\cong \pm [D_4(2)\times_{D_4(2)} D_4(2)]\cong  \ZZ_2^3\\[8pt]
|G_1|=8
\end{matrix} \qquad \qquad \qquad 
\begin{array}{r|ccc}
\text{Order } & 1 & 2 & 2 \\ W^+(E_8)\text{ class } & 1A & 2B & 2E  
\\ \# \text{ elements } &  1 & 6 & 1
\end{array}
$$

\bigskip

\noindent Symmetry group $G_0=  \D_{8}(2)\times_{D_{4}(2)} \D_{8}(2)$, order $|G_0|=16$,  with generators
$$G_0=\langle\  (1,-1),\quad (\ii,\ii), \quad (\jj,\jj)\ \rangle
$$

\bigskip

\noindent Lattice of winding momenta:
\begin{multline*}
\Gamma^{4,4}_\textnormal{w--m}=\left\{\frac{1}{\sqrt{2}}(a_1+a_2\,\ii+a_3\,\jj+a_4\,\kk;\ b_1+b_2\,\ii+b_3\,\jj+b_4\,\kk)\mid a_i,b_i\in\ZZ,\ a_i+b_i\in 2\ZZ \right\}
\end{multline*}

\bigskip

\noindent Purely left- and purely right-moving momenta:
$$\Gamma^{4,4}_\textnormal{w--m}\cap \Pi_L = \Gamma^{4,4}_\textnormal{w--m}\cap \Pi_R=\{\sqrt{2}(a+b\ii+c\jj+d\kk)\mid a,b,c,d\in\ZZ\}\cong \Lambda_{A_1^4}$$

\bigskip

\noindent Geometric description as torus model on $\RR^4/\ZZ^4$ and B-field $B=0$
$$L=
\begin{pmatrix}
1  & 0 & 0 & 0\\ 
 0 & 1 &  0 & 0 \\
 0 & 0 &1 & 0 \\
 0 &0  & 0&1 
\end{pmatrix}
\qquad 
 Q_L=\begin{pmatrix}
 1  & 0 & 0 & 0\\ 
 0 & 1 &  0 & 0 \\
 0 & 0 &1 & 0 \\
 0 &0  & 0&1
\end{pmatrix}\qquad\qquad  B= \begin{pmatrix}
0  & 0 & 0 & 0\\ 
 0 & 0 &  0 & 0 \\
 0 & 0 &0 & 0 \\
 0 &0  & 0&0\end{pmatrix}
$$

\section{Twining genera and orbifolds}\label{s:orbifolds}

In this section, we will discuss the elliptic genus of unitary $\N=(4,4)$ superconformal models \cite{EOTY}. In particular, we will focus on models with central charge $c=\tilde c=6$ whose spectrum contains a quartet of fields generating the spectral-flow isomorphism between the NS-NS and the R-R spectrum. As stressed in section \ref{s:Ramond}, the latter property is related to space-time supersymmetry for the corresponding superstring compactification. The only (known) models with these properties are non-linear sigma models with target space either $T^4$ or K3 \cite{nawe01}.  We stress that the elliptic genus can be defined for much more general conformal field theories, in particular for models with $\N=(2,2)$ superconformal symmetry; most of its properties are valid in this more general setting.

 In terms of the generators of the $\N=(4,4)$ algebra, the elliptic genus is defined as
$$ \phi(\tau,z)=\Tr_{RR}(q^{L_0-\frac{c}{24}}{\bar q}^{\tilde L_0-\frac{\tilde c}{24}} y^{J_0^3} (-1)^{F_L+F_R})\ ,\qquad q:=e^{2\pi i\tau},\quad y:=e^{2\pi iz}\ .
$$  It is invariant under deformations and gets non-vanishing contributions only from right-moving BPS states. For the the $\N=(4,4)$ superconformal field theories that we are considering, $\phi(\tau,z)$ is a weak Jacobi form of weight $0$ and index $1$ \cite{EichlerZagier}. In particular, for non-linear sigma models on  $T^4$, the elliptic genus vanishes, because the R-R BPS states form an even dimensional representation for the Clifford algebra of the right-moving fermionic zero modes $\tilde \psi^a_0$, and contributions with positive and negative $(-1)^{F_R}$ cancel each other exactly.

If the CFT has a group of symmetries $G$ preserving the $\N=(4,4)$ superconformal algebra and the spectral flow generators, then for each $g\in G$   one can define the twining genus
$$ \phi_g(\tau,z)=\Tr_{RR}(gq^{L_0-\frac{c}{24}}{\bar q}^{\tilde L_0-\frac{\tilde c}{24}} y^{J_0^3} (-1)^{F_L+F_R})\ , 
$$ which depends only on the conjugacy class of $g$ in $G$. For a non-linear sigma model on $T^4$ or K3, $\phi_g$ is a weak Jacobi form of weight $0$ and index $1$ under a suitable congruence subgroup of $SL(2,\ZZ)$, which depends on $g$.

One can also construct the orbifold of the conformal field theory by the cyclic group $\langle g\rangle\cong \ZZ_N$, by introducing the $g^i$-twisted sectors, $i=1,\ldots, N-1$, and including in the spectrum of the theory only the $g$-invariant subspace of all twisted and untwisted sectors. The orbifold is a consistent CFT, provided that the level-matching condition is satisfied, i.e. that the spin $h-\tilde h$ of the $g^i$-twisted fields takes values in $\frac{i}{N}\ZZ$, for all $i\in\ZZ$ \cite{Narain:1986qm,Frohlich:2009gb}. By construction, the orbifold is a $\N=(4,4)$ superconformal field theory with central charge $c=\tilde c=6$ and a quartet of spectral flow generators, so that, if consistent, it must be a non-linear sigma model on $T^4$ or K3 \cite{nawe01}. These two cases can be distinguished by the elliptic genus of the orbifold, which is given by the formula
$$ \phi^{orb}(\tau,z)=\frac{1}{N}\sum_{i,j=0}^{N-1} \phi_{g^i,g^j}(\tau,z)\ ,
$$ where $\phi_{g^i,g^j}$ is defined as the $g^j$-twining trace over the $g^i$-twisted sector
$$ \phi_{g^i,g^j}(\tau,z)=\Tr_{RR,\ g^i\text{-twisted}}(g^jq^{L_0-\frac{c}{24}}{\bar q}^{\tilde L_0-\frac{\tilde c}{24}} y^{J_0^3} (-1)^{F_L+F_R})\ .
$$ 
In fact, in order to distinguish between a torus and a K3 model, it is sufficient to consider the value of the elliptic genus at $z=0$ (the Witten index), which is a constant equal to the Euler number of the target space
$$ \phi(\tau,z=0)=\begin{cases}0 & \text{for a torus model,}\\ 24 & \text{for a K3 model.}\end{cases}
$$

\bigskip

In the rest of this section, we will compute the twining genera and determine the kind (either torus or K3 model) of the orbifold theory for all possible symmetries $g\in G$ of a non-linear sigma model on $T^4$.

First of all, if $g$ acts trivially on the right-moving fermions $\tilde\psi^a$, $a=1,\ldots,4$, then the twining genus $\phi_g$ vanishes for the same reasons as for the elliptic genus. Furthermore, since  the fermions are included in the spectrum of the orbifold theory, also the orbifold elliptic genus $\phi^{orb}$ vanishes. This implies that if  $\langle g\rangle$ is a finite subgroup of $U(1)^4\times U(1)^4$, then the orbifold by this group, if consistent, gives again a torus model. Therefore, it is sufficient to focus on the elements $g\in G$ with non-trivial image $g_0\in G_0$ modulo $U(1)^4\times U(1)^4$. It is convenient to label each $g_0\in G_0$ by the  $W^+(E_8)$ class of the corresponding $g'\in G_1\cong G_0/\ZZ_2$  and by the sign of the eigenvalues of the lift $g_0$ when acting on the fermions $\psi^a,\tilde\psi^a$, $a=1,\ldots,4$. For example, $+3A$ denotes an element $g_0$ which is the lift to $G_0$ of some $g'\in G_1$ in the $W^+(E_8)$ class $3A$, with eigenvalues $+\zeta_L,+\zeta_L^{-1},+\zeta_R,+\zeta_R^{-1}$ as listed in table \ref{t:classes}. For the classes acting asymmetrically between left- and right-movers, we use a prime, as in $\pm 4A'$, to denote the lift $g_0$ where the left-moving eigenvalues $\pm\zeta_L,\pm\zeta_L^{-1}$ and the right-moving ones $\pm\zeta_R,\pm\zeta_R^{-1}$ are exchanged  with respect to the ones listed in table \ref{t:classes}.

\bigskip

For a given $g\in G$, whose corresponding image $g_0\in G_0$ has eigenvalues $\zeta_L,\zeta_L^{-1},\zeta_R,\zeta_R^{-1}$,  the twining genus is given by
$$ \phi_g(\tau,z)=\phi^{gd}_g(\tau,z)\phi^{osc}_g(\tau,z)\phi^{w-m}_g(\tau)\ ,
$$ where the contribution from the $16$ R-R ground states is
$$ \phi^{gd}_g(\tau,z)=y^{-1}(1-\zeta_L y)(1-\zeta^{-1}_L y)(1-\zeta_R)(1-\zeta^{-1}_R)=2(1-\Re(\zeta_R))(y^{-1}+y-2\Re(\zeta_L))\ ,
$$ the contribution from the left-moving fermionic and bosonic oscillators is
 $$ \phi^{osc}_g(\tau,z)=\prod_{n=1}^\infty \frac{(1-\zeta_L yq^n)(1-\zeta^{-1}_L yq^n)(1-\zeta_L y^{-1}q^n)(1-\zeta^{-1}_L y^{-1}q^n)}{(1-\zeta_L q^n)^2(1-\zeta^{-1}_L q^n)^2}\ ,
$$ and the contribution from winding-momentum is
\beq\label{wmcontr} \phi_g^{w-m}(\tau)=\sum_{(\vec{\lambda}_L,0)\in (\Lambda_L)^{g_0}} q^{\frac{\vec{\lambda}_L^2}{2}}e^{2\pi i \vec{v}_L\cdot \vec{\lambda}_L}=:\Theta_{(\Lambda_L)^{g_0},v}(\tau)\ .
\ee Here, $\Lambda_L:=\Gamma^{4,4}_\textnormal{w--m}\cap \Pi_L$ and $v\equiv (\vec{v}_L,\vec{v}_R)\in (\Gamma^{4,4}_\textnormal{w--m}\otimes \RR)/\Gamma^{4,4}_\textnormal{w--m}$ determines an element of $U(1)^4_L\times U(1)^4_R$ as in eq.\eqref{U1transf}. The choice of $v$ amounts to choosing a lift $g\in G$ of $g_0\in G_0$. In the derivation of eq.\eqref{wmcontr}, we also used the fact that $\xi_g$ can be chosen so that $\xi_g(\lambda)=1$ for all $\lambda$ fixed by $g_0$. Notice that $\phi_g^{w-m}(\tau)$ is the only factor of $\phi_g$ which potentially depends on the choice of $v$. When the lattice $\Lambda_L$ contains no non-zero vectors fixed by $g_0$, we obtain $\phi_g^{w-m}(\tau)=1$ and the dependence on $v$ cancels.

 The twining genus $\phi_g$ can be written in terms of the eta function and  theta functions as
 \beq\label{twin1} \phi_g(\tau,z)=(\zeta_R+\zeta_R^{-1}-2)\frac{\vartheta_1(\tau,z)^2}{\eta(\tau)^6}\Theta_{\Lambda_L,v}(\tau)\ ,\qquad \text{for }\zeta_L= 1
\ee or
\beq\label{twin2} \phi_g(\tau,z)=(\zeta_R+\zeta_R^{-1}-2)(\zeta_L+\zeta_L^{-1}-2)\frac{\vartheta_1(\tau,z+r_L)\vartheta_1(\tau,z-r_L)}{\vartheta_1(\tau,r_L)\vartheta_1(\tau,-r_L)}\ ,\qquad \text{for }\zeta_L\neq 1
\ee
 where $r_L\in \QQ/\ZZ$ is such that $\zeta_L=e^{2\pi i r_L}$ and
$$ \vartheta_1(\tau,z)=-iq^{\frac{1}{8}}(y^{-\frac{1}{2}}-y^{\frac{1}{2}})\prod_{n=1}^\infty (1-q^n)(1-yq^n)(1-y^{-1}q^n)\ ,
\qquad \eta(\tau)=q^{\frac{1}{24}}\prod_{n=1}^\infty (1-q^n)\ .
$$
%
%
\begin{table}[!bt]
\newcolumntype{C}{>{$}c<{$}}
\begin{tabular}{CCCCCCC} g & o(g) & \dim \Hh_{U} & \dim\Hh_{g^1}+\ldots+\dim\Hh_{g^{N-1}} & \dim\Hh & \Tr_{\bf 24}(Q_g) & Q_g\\
-1A & 2&  8 & 16 & 24 & -8 & 2C\\
+3A & 3& 6 & 9+9 & 24 & -3& 3C\\
\pm 2B & 4& 6 & 4+10+4 & 24 & -4 & 4F\\
-3A & 6& 6 & 1+5+6+5+1 & 24&  -4 & 6G\\
\hline
-4A & 4& 4 & 8 +4+ 8 & 24 &0 & 4D\\
+5A & 5& 4 & 5+5+5+5 & 24& -1 & 5C\\
-3E & 6& 4 & 4+2+8+2+4 & 24 & -2 & 6L\\
\pm 6A & 6&  4 & 3+6+2+6+3 & 24 & -1 & 6M\\
\pm 4C &8&  4 & 2+3+2+6+2+3+2 & 24 &  -2 & 8H\\
-5A & 10& 4 & 1+3+1+3+4+3+1+3+1 & 24 & -2 & 10F \\
\pm 6D & 12& 4 & 1+1+2+3+1+4+1+3+2+1+1 & 24 & -2 & 12N
\end{tabular}
\caption{\small For each $W^+(E_8)$ conjugacy class of $g\in G_0$ with orbifold K3, we provide the order $o(g)$ in $G_0$, the dimension of the untwisted R-R ground sector $\Hh_U$, the dimensions of the $g$-invariant twisted sectors $\Hh_{g^k}$, $k=1,\ldots, N-1$, $N=o(g)$, the total dimension of the space $\Hh$ of R-R ground states in the orbifold (which is always $24$ for a K3 model), the trace of the quantum symmetry $Q_g$ over $\Hh$ and the $Co_0$ class of $Q_g$, as reported in \cite{Gaberdiel:2012um}. }\label{t:orbif}\end{table}
\noindent From eqs.\eqref{twin1} and \eqref{twin2}, we obtain
$$ \phi_{g^k}(\tau,0)=(1-\zeta_L^k)(1-\zeta_L^{-k})(1-\zeta_R^k)(1-\zeta_R^{-k})\ .
$$ In particular, when $\zeta_L=1$ or $\zeta_R=1$, the Witten index $\phi_g(\tau,0)$ vanishes. Using
$$ \phi_{g^i,g^j}(\tau,0)=\phi_{g^{\gcd(i,j,N)}}(\tau,0)\ ,\qquad 1\le i,j\le N\ ,
$$ where $N$ is the order of $g$, we can easily compute the Witten index of the orbifold theory. In particular, the orbifold is a torus model if $g_0$ is in one of the classes
\beq\label{torusclasses} +1A,\  \pm 2A,\ \pm 2E,\ +3E,\ +3E',\ +4A,\ +4A'\ ,
\ee while the orbifold is a K3 model when $g_0$ is in any of the other classes. In the latter case, the level-matching condition is always satisfied, while for the classes \eqref{torusclasses} it puts non-trivial constraints on the possible lifts $g\in G$ of $g_0\in G_0$.


In \cite{Gaberdiel:2012um}, a similar analysis was performed for non-linear sigma models on K3: all possible symmetries of such models were classified according to whether the corresponding orbifold is a torus or a K3 model. As shown in \cite{K3symm}, each symmetry of K3 models can be labeled by a conjugacy class in the Conway group $Co_0$, which is determined by its action on the $24$-dimensional space of R-R ground states. For each symmetry, the 'nature' of the corresponding orbifold theory only depends on its $Co_0$ class \cite{Gaberdiel:2012um}. 

We can now compare the analysis of \cite{Gaberdiel:2012um}  with the results of the present paper. More precisely, suppose that $g\in G$ is a symmetry of a torus model $\C_{T^4}$, such that the corresponding orbifold theory is a K3 model $\C'_{K3}$. This K3 model always contains a `quantum symmetry'  $Q_g$ that acts by $e^{\frac{2\pi i k}{N}}$ on the $g^k$-twisted sector.  The original CFT $\C_{T^4}$ can be re-obtained by taking the orbifold of $\C'_{K3}$ by $Q_g$.  Since we are able to compute the dimension $\dim\Hh_{g^k}=\frac{1}{N}\sum_{i=1}^N\phi_{g^k,g^i}(\tau,0)$ of the space of $g$-invariant $g^k$-twisted R-R ground states, we can compare the eigenvalues of the quantum symmetries $Q_g$ with the eigenvalues expected for a Conway class in the $24$-dimensional representation. The results, summarized in  table \ref{t:orbif}, match perfectly with the expectations from \cite{Gaberdiel:2012um}.\footnote{The classes 2C, 3C, 4F, 6G, 4D, 5C, 6L, 6M, 8G, 10F, 12N of $Co_0\cong \ZZ_2.Co_1$ in table \ref{t:orbif} map, respectively, to the classes 2A, 3C, 4C, 6C, 4B, 5C, 6E, 6F, 8E, 10D, 12I of $Co_1$ in the Atlas notation \cite{Atlas}.}

Our argument implies that there are K3 models that are orbifolds of non-linear sigma models on $T^4$ by symmetries of order $2,3,4,5,6,8,10,12$. The torus model with a symmetry of order $5$, together with its orbifold K3 model, was studied in \cite{Gaberdiel:2012um}; it turns out that this is the model of section \ref{s:A4} with symmetry group $G_0=\I_{120}\times_{\I_{120}}\bar\I_{120}$.  Furthermore, the torus model based on the lattice $\Lambda_{D_4}$ of section \ref{s:D4} was considered in \cite{GTvW}, where it was stressed that a symmetry in the class $-4A$ exists. Its orbifold is the K3 model with the `largest symmetry group' $\ZZ_2^8:M_{20}$ \cite{GTvW}.

\section{Conclusions}\label{s:conclusions}

In this paper we consider supersymmetric non-linear sigma models with target space a four dimensional torus $T^4$ and classify all possible group of symmetries of such models that preserve the $\N=(4,4)$ superconformal algebra. This results extends the analysis of \cite{K3symm} on non-linear sigma models on K3 to include all known conformal field theories with $\N=(4,4)$ superconformal symmetry at central charge $6$  and containing a quartet of spectral-flow generators. These models arise as internal CFTs in type II superstring compactifications to six dimensions that preserve half of the total space-time supersymmetries.

\medskip

One of the immediate by-products of our investigation is the description of new interesting torus models with a large amount of discrete symmetries. The new results concern mainly the models admitting left-right asymmetric symmetries with no geometric interpretation. It turns out that there are only few isolated points in the moduli space where such symmetries are realized. 

Some of the corresponding torus models are the usual suspects: the simplest example is the product of four orthogonal circles at the self-dual radius, with vanishing B-field (section \ref{s:4A1}). Another well-known example is the model with target space $\RR^4/D_4$, where $D_4$ is the root lattice of the $so(8)$ algebra (section \ref{s:D4}); the discrete symmetries and some of the orbifolds of the latter model were recently investigated in \cite{tawe11,GTvW}. The chiral algebra  in these two  theories is much larger than the one at a generic point in the moduli space $\M_{T^4}$. 

On the contrary, some of the new models considered in section \ref{s:groups} have the \emph{smallest} possible chiral algebra for a non-linear sigma model on $T^4$. These cases include, in particular, the model with a symmetry of order $5$, considered in \cite{Gaberdiel:2012um}: we discover that its group of symmetries $G_0$ is actually isomorphic to the binary icosahedral group $\I_{120}$, and that the lattice of winding-momenta is most easily described in terms of the icosian ring of unit quaternions (section \ref{s:A4}). Another interesting model has a symmetry group $G_0$ isomorphic to the binary dihedral group $\D_{24}(6)$ of order $24$ and its lattice of winding-momenta is based on the Eisenstein integers (section \ref{s:A1A1A2}).

\medskip

Our results can find some useful applications to the study of toroidal and K3 superstring compactifications. In particular, it would be interesting to understand how this investigation is related to the groups of symmetries of heterotic strings compactified on $T^4$, since the supersymmetric side of these models is analogous to the chiral half of the type II models we consider.

\medskip

In the context of the Mathieu moonshine phenomenon, the analysis of the present paper can lead to a generalization of the results of \cite{SecondQuantized}. In this work, some Siegel modular forms were constructed as the multiplicative lifts of the twisted-twining genera of generalized Mathieu moonshine \cite{Gaberdiel:2012gf}. Many of these modular forms admit an interpretation as partition functions for 1/4 BPS states in four dimensional CHL models with $16$ space-time supersymmetries. In \cite{SecondQuantized}, it was observed that there exists a particular modular transformation, induced by electric-magnetic duality in the four dimensional CHL model, that exchanges the multiplicative lifts of twining genera in two distinct K3 models, related by an orbifold. An obvious generalization of the Siegel modular forms of \cite{SecondQuantized} can be obtained by taking the multiplicative lifts of twining genera in torus models, such as the ones considered in section \ref{s:orbifolds} of the present paper. In this case, one expects the `electric-magnetic duality' to relate these forms to the multiplicative lifts of the twining genera in exceptional K3 models, i.e. the K3 models admitting a group of symmetries not in $M_{24}$ \cite{Gaberdiel:2012um}. It would be very interesting to investigate in detail this instance of electric-magnetic duality and to understand its physical meaning.

A somehow related open question concerns the existence of any kind of moonshine phenomenon for non-linear sigma models on $T^4$. By analogy with the K3 case, one would expect the corresponding twining genera to reproduce some of the functions computed in section \ref{s:orbifolds}.\footnote{We thank M.~R.~Gaberdiel for discussions about this possibility.}

\medskip

The symmetries considered in the present paper can be described in terms of topological defects preserving a small $\N=(4,4)$ superconformal algebra. The orbifold procedure relating the corresponding torus models to K3 has also a natural interpretation in terms of defects \cite{Frohlich:2006ch}. At the moment, only the topological defects of $d$-dimensional torus models that preserve the $u(1)^d\oplus u(1)^d$ current algebra have been classified \cite{Bachas:2012bj}. In the same spirit,  it is natural to ask whether the classification of this paper includes all possible defects preserving the $\N=(4,4)$ algebra or if more possibilities exist. 
  In the latter case, one might obtain new explicit descriptions of K3 models arising from the associated generalized orbifolds of torus models \cite{Frohlich:2009gb,Carqueville:2012dk,BruCarqPle}.

\acknowledgments

I would like to thank Stefan Fredenhagen, Matthias R. Gaberdiel and Anne Taormina for useful comments on the manuscript.

\appendix

\section{Discrete subgroups of ${Spin(4)}$ and $SO(4)$}\label{s:discrSpin}

For the classification of the  discrete subgroups of $\Spin(4)\cong SU(2)\times SU(2)$ and $SO(4)\cong \Spin(4)/(-1,-1)$ we follow closely the treatment of \cite{ConwaySmith}.  We recall that $SU(2)$ can be described  as the group of unit quaternions
$$ SU(2)= \{ a+b\ii+c\jj+d\kk\in \HH\mid a^2+b^2+c^2+d^2=1\}\ ,
$$ and its discrete subgroups are as in table \ref{t:discrsu2}.
\begin{table}[t]
\begin{center}
\begin{tabular}{cclccc}
 $\G\subset SU(2)$& $|\G|$ & Name  & Image in $SO(3)$ & Generators \\
 \hline
$\C_n\begin{array}{l}
(\equiv 1C_n)\\
(\equiv 2C_{n/2})
\end{array}$& $\begin{matrix}
\text{odd }n\\
\text{even }n
\end{matrix}$ & cyclic & $\begin{matrix}C_n\\
C_{n/2}\end{matrix}$ & $\eei_{1/n}$\\
$ \D_{4n}(n)\equiv 2D_{2n}(n)$ &$4n$ & (binary) dihedral & $D_{2n}(n)$ & $\eei_{1/2n},\jj$\\
$\T_{24}\equiv 2T_{12}$ &24 & (binary) tetrahedral & $T_{12}$  & $\ii,\om$\\
$\calO_{48}\equiv 2O_{24}$ &48 & (binary) octahedral & $O_{24}$ & $\iiO,\om$ \\
$\I_{120}\equiv 2I_{60}$&120 & (binary) icosahedral & $I_{60}$ & $\iiI,\om$
\end{tabular}
\end{center}
\caption{\small The discrete subgroups $\G$ of $SU(2)$, their order $|\G|$, their name, the image in $SO(3)=SU(2)/(-1)$ and a set of generators in the form of quaternions. Here, $
\eei_r=\exp(2\pi \ii r)$, $\om=\frac{-1+\ii+\jj+\kk}{2}$, $\iiO=\frac{\jj+\kk}{\sqrt{2}}$,  $\iiI=\frac{1}{2}(\ii+\tfrac{-\sqrt{5}-1}{2}\jj+\tfrac{\sqrt{5}-1}{2}\kk)$.
}\label{t:discrsu2}\end{table}
 Notice that $C_n\subset SO(3)$, with odd $n$, admits both  a double and a single cover in $SU(2)$, namely $\C_{n}$ and $\C_{2n}$, while all the other subgroups of $SO(3)$ admit only a double covering.

The simplest subgroups of $SU(2)\times SU(2)\cong \Spin(4)$ are just direct products $\L\times \R$, where $\L,\R$ are discrete subgroups of $SU(2)$. 
We denote by $[l,r]\equiv [-l,-r]$ the image of an element $(l,r)\in \L\times \R$ under the covering map  $ \Spin(4)\to (SU(2)\times SU(2))/(-1,-1)\cong SO(4)$. In particular,  $[l,r]\equiv [-l,-r]\in SO(4)$ acts on $\RR^4\cong \HH$ by
$$ [l,r]:q\mapsto  l q \bar r\ ,\qquad q\in \HH\ ,
$$ where $\bar r$ denotes the complex conjugate of $r$.

A generic discrete $\G\subset SU(2)\times SU(2)$ is a subgroup of a direct product group $\G\subseteq \L\times \R$. Any such subgroup $\G$ can be defined in terms of two surjective homomorphisms
$$ \alpha:\L \to \F\ ,\qquad \qquad
\beta:\R \to \F\ ,
$$
from $\L$ and $\R$ onto the same abstract group $\F$. More precisely, $\G$ is the group of pairs $(l,r)\in \L\times\R$ such that $l$ and $r$ have the same image in $\F$
$$ \G=\{(l,r)\in \L\times \R \mid \alpha(l)=\beta(r) \}\ ,
$$ for some suitable $\alpha,\beta$. In particular, $\L_0:=\ker\alpha$ and $\R_0:=\ker\beta$ are (isomorphic to) the normal subgroups of $\G$ generated by elements of the form $(l,1)\in \G$ and $(1,r)\in \G$, respectively, so that
$$ \L_0\times \R_0 \subseteq \G\subseteq \L\times \R\ .
$$ 
For example, when $|\F|=1$, we have $\L_0=\L$, $\R_0=\R$ and $\G$ is simply the direct product $\L\times \R$. At the opposite extreme, when $\L=\F=\R$ and $\alpha,\beta$ are the identity maps, then $\L_0=1=\R_0$ and $\G$ is also isomorphic to $\F$. The groups of purely geometric symmetries of torus models are of this kind.

In most cases, giving the groups $\L$, $\R$ and $\F$ is sufficient to determine  the homomorphisms $\alpha$ and $\beta$ uniquely (up to conjugation of $\L$ and $\R$ in $SU(2)$ and up to automorphisms of $\F$). This justifies the notation $\L\times_{\F} \R$ for the corresponding group $\G$. When there are inequivalent choices for $\alpha,\beta$, we denote the `most obvious' group by $\L\times_{\F} \R$ and the other ones by some decorations, such as $\L\times_{\F} \bar\R$. The precise structure of each group will be clear from its list of generators. For example, $\I_{120}\times_{\I_{120}}\I_{120}$ denotes the group where both $\alpha,\beta$ are the identity map, while $\I_{120}\times_{\I_{120}}\bar\I_{120}$ denotes the group where $\alpha$ is the identity map and $\beta$ is a non-trivial outer automorphism of $\I_{120}$. The two groups are actually isomorphic as abstract groups (and isomorphic to $\I_{120}$), but they are not conjugated within $SU(2)\times SU(2)$.

For the subgroups of $SO(4)\cong (SU(2)\times SU(2))/(-1,-1)$ and $PSO(4)\cong SU(2)/(-1)\allowbreak \times\allowbreak SU(2)/(-1)$, the treatment is analogous. We denote the subgroups of $PSO(4)$ by ${[L\times_F R]}$, where $L,R,L_0,R_0 \subset SU(2)/(-1)$ and $L/L_0\cong F\cong R/R_0$. Any such group has a double covering in $SO(4)$, denoted by $\pm[L\times_F R]$, which contains the element $[1,-1]\equiv [-1,1]\in SO(4)$. In particular, the elements of $\pm[L\times_F R]$ always come in pairs $[l,r]$, $[l,-r]$, with the same image in  $[L\times_F R]$. In some cases, there exists an index two subgroup of $\pm[L\times_F R]$ which contains only one element (either $[l,r]$ or $[l,-r]$) for each pair; we denote this group by $+[L\times_F R]$. Notice that $+[L\times_F R]\subset SO(4)$ does not contain the element $[1,-1]\in SO(4)$ and is isomorphic, as an abstract group, to $[L\times_F R]\subset PSO(4)$. This case can only occur when both $L_0$ and $R_0$ are cyclic groups of odd order.

For each finite $G\subset SO(4)$ it is easy to find its double covering $\G \subset \Spin(4)$ (see table \ref{t:doublecover}). In the table, $\L,\R,\L_0,\R_0\subset SU(2)$ denote the double coverings of $L,R,L_0,R_0\subset SO(3)\cong SU(2)/(-1)$, respectively, so that $\L/\L_0\cong F\cong \R/\R_0$. Moreover, $\F\cong \ZZ_2.F$ denotes a central extension of $F$. More precisely, when the group $+[L\times_{F} R]$ is defined, there exist index $2$ subgroups $\L_0'\subset \L_0$ and $\R_0'\subset\R_0$ that are 1-to-1 covering of $L_0$ and $R_0$, respectively, i.e. $\L_0'\stackrel{\cong}{\rightarrow} L_0$ and $\R_0'\stackrel{\cong}{\rightarrow} R_0$. In this case, $\F$ is given by the quotient $\L/\L_0'\cong \F\cong \R/\R_0'$.
\begin{table}[t]\begin{center}
\begin{tabular}{cccc}
$G\subset SO(4)$ & $|G|$ & Lift $\G\subset \Spin(4)$ & $|\G|$\\
\hline\\[-10pt]
$\pm[L\times_{F} R]$ & $2\frac{|L|\times |R|}{|F|}$ & $\L\times_{F} \R$ & $\frac{|\L|\times |\R|}{|F|}$\\[5pt]
$+[L\times_{F} R]$ & $\tfrac{|L|\times |R|}{|F|}$ & $\L\times_{\F} \R$ & $\frac{|\L|\times |\R|}{|\F|}$
\end{tabular}\end{center}
\caption{\small Discrete subgroups of $SO(4)$ and their double covering in $\Spin(4)$. }\label{t:doublecover}
\end{table}

\section{Lattices}\label{s:lattices}

In this appendix, we review some general properties of even lattices and their automorphisms, in particular   the `gluing' construction (see \cite{Conway} and \cite{Nikulin} for more details).

For any lattice $\Lambda\subset \RR^n$, we denote by $\Lambda^*$ its dual and by $\Lambda(-1)$ the lattice obtained by changing the sign of the quadratic form.  We define the determinant of $\Lambda$ to be the determinant $|\det Q|$ of the Gram matrix $Q_{ij}:=\lambda_i\cdot \lambda_j$, for some basis $\lambda_1,\ldots,\lambda_r$ of generators of $\Lambda$. The determinant is independent of the choice of the basis. From now on, we assume that $\Lambda$ is non-degenerate, i.e. $|\det Q|\neq 0$, and we identify $\Lambda\otimes\RR$ with $\Lambda^*\otimes\RR$ through the induced metric.  If $\Lambda$ is integral, and in particular if it is even, then $\Lambda \subset \Lambda^*$ and we can define the \emph{discriminant group} $\Lambda^*/\Lambda$. This is a finite abelian group of order $|\det Q|$. 

Let us assume that $\Lambda$ is even. The quadratic form $\Lambda\to 2\ZZ$ on $\Lambda$ induces a quadratic form $\Lambda^*\to \QQ$ on the dual lattice $\Lambda^*$ and the discriminant  form
$$ q_{\Lambda}:\Lambda^*/\Lambda\to \QQ/2\ZZ\ ,
$$ on the discriminant group $\Lambda^*/\Lambda$. 

A sublattice $\Lambda'$ of $\Lambda$ is called primitive if $\Lambda/\Lambda'$ is a free group, i.e. if $\Lambda\cap(\Lambda'\otimes \QQ)=\Lambda'$. Let $\Lambda$ be a primitive sublattice of an even unimodular lattice $\Gamma$ and let  $\Lambda^\perp$ be its orthogonal complement $\Lambda^\perp=\Gamma\cap (\Lambda\otimes\RR)^\perp$. Then, there exists an isomorphism of discriminant groups
$$ \gamma: \Lambda^*/\Lambda\stackrel{\cong}{\longrightarrow} (\Lambda^\perp)^*/\Lambda^\perp\ ,
$$ that inverts the discriminant quadratic form
\beq\label{invert} q_{\Lambda^\perp}\circ \gamma = - q_{\Lambda}\mod 2\ZZ\ ,
\ee and such that
\beq\label{unimodu} \Gamma = \{(v,w)\in \Lambda^*\oplus_\perp (\Lambda^\perp)^*\mid [w]=\gamma([v])\}
\ee where for any $x\in \Lambda^*$ (respectively, $x\in (\Lambda^\perp)^*$) $[x]$ denotes the class of $x$ in $\Lambda^*/\Lambda$ (respectively, in $(\Lambda^\perp)^*/\Lambda^\perp$). Vice versa, for any pair of even lattices $\Lambda$ and $\Lambda^\perp$ with an isomorphism $\gamma: \Lambda^*/\Lambda\stackrel{\cong}{\longrightarrow} (\Lambda^\perp)^*/\Lambda^\perp$ satisfying \eqref{invert}, the lattice  $\Gamma$ defined by eq.\eqref{unimodu} is even unimodular. This is called the `gluing' construction. Clearly, a necessary condition for the isomorphism $\gamma$ to exist is that the discriminant groups have the same order, i.e. the two lattices have the same determinant.

Any automorphism $g\in O(\Lambda)$ induces an automorphism of the dual $\Lambda^*$ and of the discriminant group, preserving the discriminant form $q_\Lambda$. We denote by $O_0(\Lambda)$ the subgroup of automorphisms acting trivially on the discriminant group and by $SO(\Lambda)$ and $SO_0(\Lambda)$ the corresponding subgroups of orientation-preserving automorphisms. If $\Lambda$ and $\Lambda^\perp$ are two primitive mutually orthogonal sublattices of the even unimodular lattice $\Gamma$, as above, then any element $g\in O_0(\Lambda)$ extends to an automorphism	of $\Gamma$ \cite{Nikulin}. Indeed, $g$ obviously extends to an automorphism of $\Lambda^*\oplus_\perp (\Lambda^\perp)^*$ by $(v,w)\mapsto (g(v),w)$ and since $[v]=[g(v)]$ for all $v\in \Lambda^*$ this automorphism preserves the sublattice $\Gamma$.

\end{document}